\documentclass[aps,pra,twocolumn,nofootinbib,floatfix,superscriptaddress]{revtex4-2}
\usepackage{amsmath,mleftright,mathtools}
\usepackage{bm}
\usepackage{stix,microtype}
\usepackage{graphicx}
\usepackage{xspace}
\usepackage{units}
\usepackage[colorlinks,linkcolor=blue,citecolor=blue,urlcolor=blue]{hyperref}
\usepackage[dvipsnames]{xcolor}
\usepackage{cleveref}
\usepackage{array}   % for \newcolumntype macro
\newcolumntype{L}{>{$}c<{$}} % math-mode version of "l" column type
\newcolumntype{C}{>{$}c<{$}} % math-mode version of "l" column type
\usepackage{dcolumn}
\newcommand{\hphi}{\hat{\phi}}

\newcommand{\s}{\sigma}
\graphicspath{{./Figs/}}
\mathchardef\mhyphen="2D

\newcommand{\be}{\begin{equation}}
\newcommand{\ee}{\end{equation}}

\newcommand{\mv}{\mathbfit{v}}

\newcommand{\mh}{\mathbfit{h}}

\newcommand{\mO}{\bm{\Omega}}
\newcommand{\mD}{\mathbfit{D}}

\newcommand{\mPsi}{\bm{\Psi}}

\newcommand{\mSgm}{\mathbfit{\Sigma}}

\newcommand{\mPi}{\bm{\Pi}}

\newcommand{\cG}{\mathcal{G}}

\newcommand{\mcG}{\bm{\mathcal{G}}}
\newcommand{\mcA}{\bm{\mathcal{A}}}
\newcommand{\mI}{\mathbfit{I}}

\newcommand{\mP}{\mathbfit{P}}
\newcommand{\mrho}{\bm{\rho}}

\newcommand{\bgg}{\bm{\gamma}}
\newcommand{\mchi}{\bm{\mathit{\chi}}}

\def\bgm{\mbox{\boldmath $\mu$}}

\def\fbgm{\mbox{\scalebox{.7}{$\scriptscriptstyle \bgm$}}}

\def\bga{\mbox{\boldmath $\alpha$}}
\def\bgd{\mbox{\boldmath $\delta$}}
\def\bgg{\mbox{\boldmath $\gamma$}}
\def\bcallg{\mbox{\boldmath $g$}}

\def\w{\omega}

\keywords{Nonequilibrium Green's function theory, generalized Kadanoff-Baym Ansatz, excited states}
\begin{document}
\title{Interacting electrons and bosons in the doubly screened $G\widetilde{W}$ approximation:
A time-linear scaling method for first-principles simulations}
\author{Y. Pavlyukh}
\affiliation{Dipartimento di Fisica, Universit{\`a} di Roma Tor Vergata, Via della Ricerca Scientifica 1,
00133 Rome, Italy}
\affiliation{Department of Theoretical Physics,
  Faculty of Fundamental Problems of Technology,
  Wroc{\l}aw University of Science and Technology,
50-370 Wroc{\l}aw, Poland}
%\email{yaroslav.pavlyukh@gmail.com}
\author{E. Perfetto}
\affiliation{Dipartimento di Fisica, Universit{\`a} di Roma Tor Vergata, Via della Ricerca Scientifica 1,
00133 Rome, Italy}
\affiliation{INFN, Sezione di Roma Tor Vergata, Via della Ricerca Scientifica 1, 00133 Rome, Italy}
% \author{Daniel Karlsson}
% \affiliation{Department of Physics, Nanoscience Center P.O.Box 35
% FI-40014 University of Jyv\"{a}skyl\"{a}, Finland}
% \author{Robert van Leeuwen}
% \affiliation{Department of Physics, Nanoscience Center P.O.Box 35
% FI-40014 University of Jyv\"{a}skyl\"{a}, Finland}
\author{G. Stefanucci}
\affiliation{Dipartimento di Fisica, Universit{\`a} di Roma Tor Vergata, Via della Ricerca Scientifica 1,
00133 Rome, Italy}
\affiliation{INFN, Sezione di Roma Tor Vergata, Via della Ricerca Scientifica 1, 00133 Rome, Italy}
\date{\today}
\begin{abstract}  
We augment the time-linear formulation of the Kadanoff-Baym 
equations for systems of interacting electrons and quantized phonons 
or photons with the $G\widetilde{W}$ approximation, 
the Coulomb interaction $\widetilde{W}$ being dynamically screened by both electron-hole 
pairs {\em and} bosonic particles. We also show how to combine 
different approximations to include simultaneously multiple correlation effects 
in the dynamics. The final outcome is a versatile framework
comprising $2^{12}$ distinct diagrammatic methods, each scaling 
linearly in time and preserving all fundamental conservation laws.
The dramatic improvement over current state-of-the-art 
approximations brought about by $G\widetilde{W}$  is demonstrated in a study of the 
correlation-induced charge migration of the  glycine
molecule in an optical cavity.

\end{abstract}
\maketitle
%===== ===== ===== ===== ===== =====   I   ===== ===== ===== ===== ===== =====
{\em Introduction:}                             
%===== ===== ===== ===== ===== ===== ===== ===== ===== ===== ===== ===== =====
% The  dynamics of correlated electrons and bosons underlies the electronic, 
% magnetic and optical properties of any quantum systems, from simple 
% molecules to bulk materials. 
After Feynman's visionary idea in 1949~\cite{feynman_space-time_1949}
the Green's function (GF) diagrammatic theory has developed into a
powerful and versatile approach in nearly every field of theoretical
physics. In condensed matter
theory~\cite{abrikosov_methods_1975,mattuck_guide_1992,fetter_quantum_2003,gross_many-particle_1991}
efforts toward the nonequilibrium extension of the formalism
(NEGF)~\cite{konstantinov_diagram_1961,keldysh_diagram_1965}
culminated in the so-called Kadanoff-Baym equations
(KBE)~\cite{kadanoff_quantum_1962,stefanucci_nonequilibrium_2013}.
The KBE govern the dynamics of correlated electrons and bosons and
give access to the electronic, magnetic and optical properties of any
quantum system, from simple molecules to bulk materials. As for any
exact reformulation of the many-body Schr\"odinger equation the
applicability of the KBE relies on accurate approximations and
efficient implementation
schemes~\cite{stan_time_2009,balzer_nonequilibrium_2013-1,schuler_time-dependent_2016,schuler_nessi_2020}.

In Ref.~\cite{pavlyukh_time-linear_2021} we built on the Generalized 
Kadanoff-Baym Ansatz (GKBA) for 
electrons~\cite{lipavsky_generalized_1986} and 
bosons~\cite{karlsson_fast_2021} and on the time-linear formulation 
of the GKBA-KBE with electron-electron ($e$-$e$)~\cite{schlunzen_achieving_2020,joost_g1-g2_2020}  
and electron-boson ($e$-$b$)~\cite{karlsson_fast_2021} interactions
to map a broad class of NEGF approximations onto a coupled system of 
ordinary differential equations (ODE). Available methods to treat 
$e$-$e$ correlations include $GW$~\cite{perfetto_real-time_2022}, $T$-matrix (either without or 
with exchange) and Faddeev~\cite{pavlyukh_photoinduced_2021}  while 
$e$-$b$ correlations are described by Ehrenfest and 
second-order diagrams in the $e$-$b$ 
coupling~\cite{frederiksen_inelastic_2007,cannuccia_effect_2011,pavlyukh_time-linearII_2021, rizzi_electron-phonon_2016}.
Every method in 
this NEGF toolbox guarantees the fulfillment of all fundamental 
conservation 
laws~\cite{baym_conservation_1961,baym_self-consistent_1962,stefanucci_nonequilibrium_2013}.

In this work we present a substantial advance in the 
treatment of correlations, requiring no extra computational cost and 
preserving all conserving properties. Specifically we include the 
effects of dynamical screening due to {\em both} $e$-$e$  and $e$-$b$ 
interactions
($G\widetilde{W}$ approximation)~\cite{van_leeuwen_first-principles_2004,andreoni_non-equilibrium_2020}. 
The  $G\widetilde{W}$  extention opens the door to 
a wealth of phenomena
ranging from carrier relaxation~\cite{sangalli_ultra-fast_2015,molina-sanchez_ab_2017}
and exciton recombination~\cite{selig_excitonic_2016,trovatello_strongly_2020} 
to molecular charge migration and transfer in optical or plasmonic 
cavities~\cite{flick_ab_2018,ojambati_quantum_2019,schafer_modification_2019,li_cavity_2021}.
We further show how to combine 
different  methods without incurring any double counting. The final 
outcome is a NEGF toolbox that can be used to 
investigate the 
correlated dynamics of electrons and bosons in 
$2^{12}$ distinct diagrammatic approximations.
Real-time simulations of the correlation-induced charge migration of the glycine molecule  
in an optical (or plasmonic) cavity demonstrates the superiority of 
the $G\widetilde{W}$ method over other approximations.
% 
% To exemplifies the impact of the 
% $G\tilde{W}$ method we simulate a {\em gedanken} 
% experiment where a molecule in an optical (or plasmonic) cavity 
% couples to a cavity-photon after photoionization.

%===== ===== ===== ===== ===== =====   I   ===== ===== ===== ===== ===== =====
{\em Preliminaries:}                             
%===== ===== ===== ===== ===== ===== ===== ===== ===== ===== ===== ===== =====
We consider a system of electrons with one-particle time-dependent 
Hamiltonian $h_{ij}(t)$ and $e$-$e$ interaction $v_{ijmn}$ (Latin indices 
$i,j,\ldots$ etc. specify the spin-orbitals of an orthonormal basis) 
coupled linearly to the displacement $\hphi_{\fbgm,1}\equiv \hat{x}_{\fbgm}=(\hat{a}^{\dag}_{\fbgm}+\hat{a}_{\fbgm})/\sqrt{2}$
and momentum 
$\hphi_{\fbgm,2}=\hat{p}_{\fbgm}=i(\hat{a}^{\dag}_{\fbgm}-\hat{a}_{\fbgm})/\sqrt{2}$ 
of a set of bosonic modes of frequency $\w_{\fbgm}$. Introducing the Greek 
index $\mu=(\bgm,\xi)$ with $\xi=1,2$, we denote by $g_{\mu,ij}$ the 
interaction strength of the $e$-$b$ coupling. The 
equation of motion (EOM) for the one-electron density matrix 
$\rho^{<}_{ij}(t)\equiv\langle \hat{d}_{j}^\dagger(t)\hat{d}_{i}(t)\rangle$ 
[with $\hat{d}^{(\dag)}$'s the electronic annihilation (creation) 
operators] and one-boson density matrix 
$\gamma_{\mu\nu}^{<}(t)\equiv\langle \Delta \hphi_{\nu}(t)\Delta 
\hphi_{\mu}(t)\rangle$ [with $\Delta \hphi_{\nu}\equiv 
\hphi_{\nu}-\langle\hphi_{\nu}\rangle$ the bosonic fluctuation 
operator] reads~\cite{karlsson_fast_2021}
\begin{subequations}
\begin{align}
  i\frac{d}{dt}\rho^<(t)&=\big[h^{e}(t),\rho^<(t) \big] 
  -i\left(I^{e}(t)+I^{e\,\dagger}(t)\right),
  \label{eq:eomrho:e}
\\
  i\frac{d}{dt}\bgg^<(t)&= \big[\mh^{b}(t),\bgg^<(t) \big]
  +i\left(\mI^{b}(t)+\mI^{b\dagger}(t)\right),
  \label{eq:eomrho:b}
\end{align}
\label{1PEOM}%
\end{subequations}
where $h^{e}_{ij}(t)=h_{ij}(t)+\sum_{mn}[v_{imnj}(t)-v_{imjn}(t)]\rho^{<}_{nm}(t)+
\sum_{\mu}g_{\mu,ij}(t)\phi_{\mu}(t)$ is the mean-field electronic 
Hamiltonian [$\phi_{\mu}=\langle\hphi_{\mu}\rangle$ for brevity] 
whereas $\mh^{b}(t)=2\bga\mO(t)$, with $\alpha_{\mu\mu'}\equiv \delta_{\fbgm\fbgm'}
  \begin{pmatrix}
  0  & i \\ -i & 0
  \end{pmatrix}_{\xi\xi'}$ and 
$\Omega_{\mu\mu'}(t)\equiv \frac{1}{2}\delta_{\mu\nu}\w_{\fbgm}(t)$, 
is the free-boson Hamiltonian. 
To distinguish matrices in the one-electron space from matrices in the one-boson
space we use boldface for the latters. The time-dependence of the 
$e$-$e$ coupling $v_{ijmn}(t)$ and $e$-$b$ coupling $g_{\mu,ij}(t)$
could be due to the adiabatic switching protocol adopted to
generate a correlated initial state~\cite{karlsson_generalized_2018}, whereas the time-dependence of 
the one-particle Hamiltonian $h_{ij}(t)$ and bosonic frequencies $\w_{\fbgm}(t)$ 
could be due to some external field, e.g., laser 
fields~\cite{bostrom_charge_2018,perfetto_ultrafast_2018}, phonon
drivings~\cite{murakami_nonequilibrium_2017}, etc. As the mean-field Hamiltonian $h^{e}$ depends on 
$\phi_{\mu}(t)$ the EOM (\ref{1PEOM}) must be complemented with the Ehrenfest 
EOM for the displacements and momenta of the bosonic modes, see 
below.
% 
% $i\frac{d}{dt}\phi_\mu(t) =\sum\nolimits_{\nu} h^{b}_{\mu\nu}(t)   \phi_\nu(t) 
% +\sum\nolimits_{ij}\bar{g}_{\mu,ij} \rho_{ji}^{<}(t)$, where we have 
% defined $\bar{g}_{\mu.ij}\equiv 
% \sum_{\nu}\alpha_{\mu\nu}g_{\nu,ij}$. 

The collision integrals $I^{e}$ and $\mI^{b}$ accounts for all 
effects beyond mean-field. They can be written  in terms of 
two high-order GFs
according to~\cite{karlsson_fast_2021} $I^{e}_{lj}=i\sum_{\mu,i} g_{\mu,li}\cG^{b}_{\mu,ij}
-i\sum_{imn} v_{lnmi} \cG^{e}_{imjn}$ and 
$I^{b}_{\mu\nu}=-i\sum_{\nu,mn} \alpha_{\mu\nu} g_{\nu,mn}\cG^{b}_{\nu,nm}$, where
\begin{align}
\cG^{e}_{imjn}(t)&=-\langle 
\hat{d}^{\dag}_{n}(t)\hat{d}^{\dag}_{j}(t)\hat{d}_{i}(t)\hat{d}_{m}(t)\rangle_{c},
\label{Gedef}
\\
\cG^{b}_{\mu,ij}(t)&=\langle\hat{d}^{\dag}_{j}(t)\hat{d}_{i}(t)\hat{\phi}_{\mu}(t)\rangle_{c}.
\label{Gbdef}
\end{align}
The subscript ``$c$'' in the averages signifies that only the correlated part must be
retained. The EOM (\ref{1PEOM}) fulfill all fundamental conservation 
laws if  
$\cG^{e}$ and $\cG^{b}$ are obtained from the functional derivatives of the 
correlated part $\Phi_{c}$ of the Baym functional~\cite{baym_self-consistent_1962}  
with respect to the $e$-$e$ and $e$-$b$ 
coupling respectively, i.e.,
\begin{subequations}
\begin{align}
  \cG^{e}_{imjn}(t)&=i  \frac{\delta \Phi_{c}}{\delta 
  v_{jnmi}(t)}+i  \frac{\delta \Phi_{c}}{\delta 
  v_{njim}(t)},
  \\
  \cG^{b}_{\mu,ij}(t)&= \frac{1}{i} \,\frac{\delta \Phi_{c}}{\delta 
  g_{\mu,ji}(t)}.
  \label{gbfuncder}
\end{align}
\label{hogs}
\end{subequations}

In Ref.~\cite{karlsson_fast_2021} we have considered the correlated 
functional
$\Phi_{c}=$\raisebox{-6pt}{\includegraphics{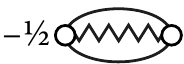}} -- full lines 
represent electronic GFs $G$, zig-zag lines  
bosonic GFs $D$ and empty circles  the $e$-$b$ coupling 
$g$.
The mathematical expression of the considered functional reads (time integrals are over the Keldysh contour)
\begin{align}
\Phi_{c}=-\frac{1}{2}\int \!d\bar{t}d\bar{t}'\,{\rm Tr}[\bcallg^{\dag}(\bar{t})\mD(\bar{t},\bar{t}')
\bcallg(\bar{t}')\mchi^{0}(\bar{t}',\bar{t})],
\label{GDPhi}
\end{align}
where we have defined the matrix $\bcallg$ with elements 
$g_{\mu\nu}=g_{\mu,\begin{subarray}{c}j\\i\end{subarray}}=g_{\mu,ij}$
(hence the second Greek-index $\nu=
\bigl(\begin{smallmatrix}
  %\begin{pmatrix}
  j\\i
 % \end{pmatrix}
  \end{smallmatrix}\bigr)$ labels a pair of electronic 
indices) and the electronic response function 
$\chi^{0}_{\mu\nu}(t',t)=
\chi^{0}_{\begin{subarray}{c}qj\\ si\end{subarray}}(t',t)\equiv
-iG_{q j} (t',t)G_{is}(t,t')$. 
Consistently with our notation,  matrices with Greek indices are represented by
boldface letters.
Through Eqs.~(\ref{hogs}) one obtains  
$\cG^{e}=0$ and $\mcG^{b}(t)=i\int d\bar{t} \mD(t,\bar{t})\bcallg(\bar{t})
  \mchi^{0}(\bar{t},t^{+})$. Implementing the  
GKBA for electrons and 
bosons~\cite{lipavsky_generalized_1986,karlsson_fast_2021}, 
\begin{align}
  G^{\lessgtr}(t,t')&=-G^{R}(t,t')\rho^{\lessgtr}(t')+\rho^{\lessgtr}(t)G^{A}(t,t'),\label{eq:e:gkba}\\
  \mD^{\lessgtr}(t,t')&=\mD^{R}(t,t')\bga\bgg^{\lessgtr}(t')-\bgg^{\lessgtr}(t)\bga \mD^{A}(t,t'),\label{eq:b:gkba}
\end{align}
one can
show that $\mcG^{b}$ satisfies a first-order 
ODE~\cite{karlsson_fast_2021} whose
coefficients are given by
simple functionals of
the density matrices $\rho^{<}$, $\rho^{>}\equiv\rho^{<}-1$ and 
$\bgg^<$, $\bgg^>\equiv \bgg^<+\bga$. This is pivotal  for 
constructing a time-linear scheme. The resulting GKBA+ODE are equivalent to  the 
original KBE --- in the GKBA framework --- with  electronic 
self-energy in the $GD$ 
approximation~\cite{fan_temperature_1951,murakami_multiple_2016,pavlyukh_time-linearII_2021} and  bosonic
self-energy proportional to $\mchi^{0}$.
The feedback of electrons (bosons) 
on the bosonic (electronic) subsystem underlies  the 
fulfillment of all conservation laws.

{\em The doubly screened $G\widetilde{W}$ method:}
The functional $\Phi_{c}$ in Eq.~(\ref{GDPhi}) is independent of 
the $e$-$e$ interaction; hence electronic screening of the 
$e$-$b$ coupling is not accounted for. This is a severe 
drawback for extended systems~\cite{giustino_electron-phonon_2007,marini_many-body_2015}.
State-of-the-art calculations of
electronic 
life-times~\cite{restrepo_first-principles_2009}, 
polaron dispersions~\cite{verdi_origin_2017} 
and carrier dynamics~\cite{molina-sanchez_ab_2017}
are indeed performed with a {\em statically} screened electron-phonon 
coupling~\cite{mahan_many-particle_2000,giustino_electron-phonon_2017,caruso_nonadiabatic_2017}. 
Formally, static screening does not involve any generalization of the $GD$ 
equations: it is sufficient to replace one of the $\bcallg$'s in Eq.~(\ref{GDPhi}) 
with $\bcallg^{s}=\bcallg(1+ \mchi^{s}\mv)$, where 
$v_{\begin{subarray}{c}im\\ nj\end{subarray}}\equiv v_{ijmn}$ and 
$\mchi^{s}$ is the random phase approximation (RPA) response function, 
$\mchi=\mchi^{0}+\mchi\mv\mchi^{0}$, evaluated in equilibrium and at zero frequency.
Although $\bcallg^{s}$ is  an 
improvement over the bare $\bcallg$,  
retardation effects and nonequilibrium corrections are still lacking.
In the following we show that a time-linear GKBA+ODE 
scheme can be formulated for the two-times {\em dynamically} 
screened  coupling
$\bcallg^{d}= \bcallg(1+ \mchi\mv)$. 

\begin{figure}[tbp]
\centering  \includegraphics[width=0.99\columnwidth]{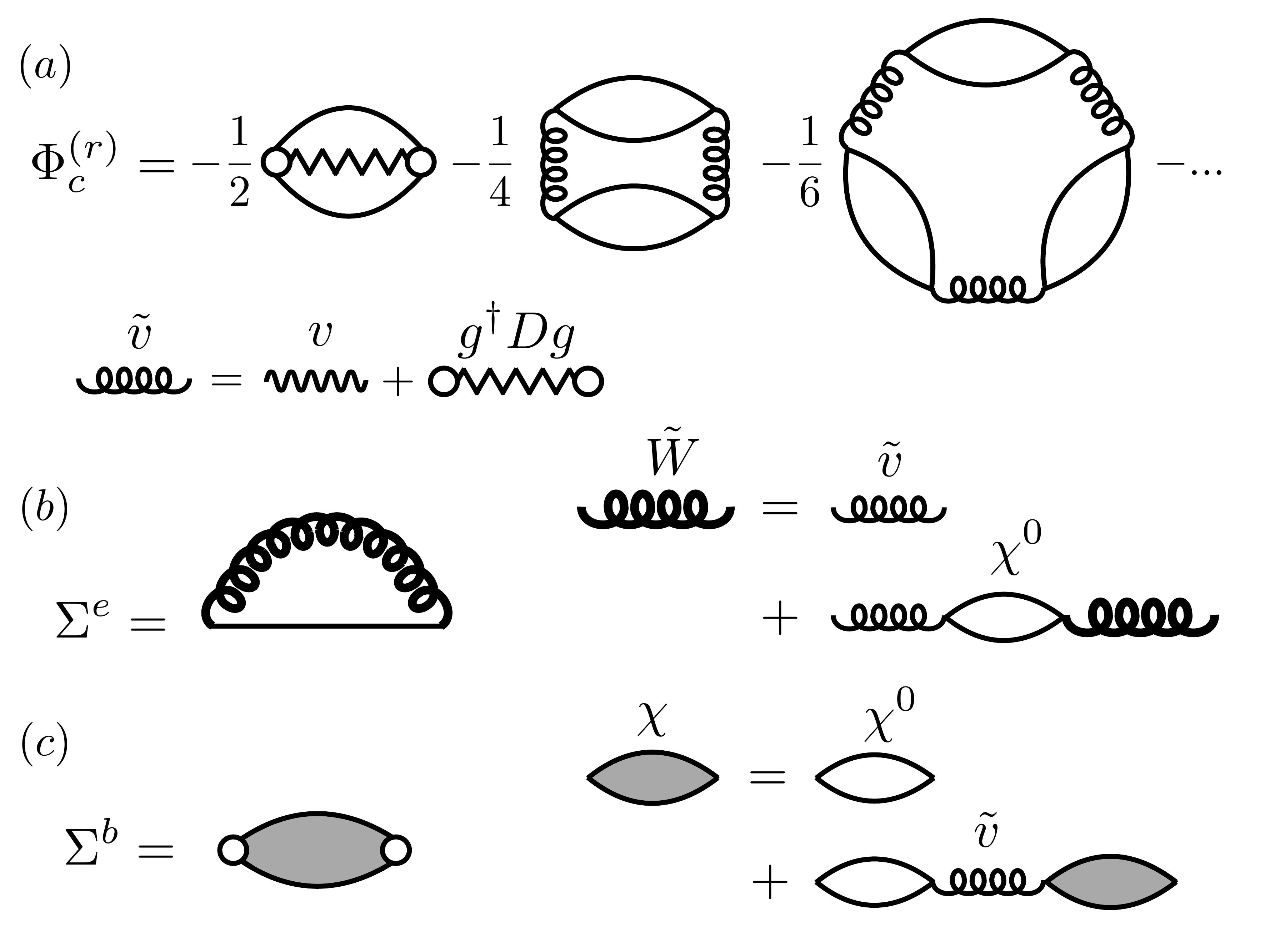}
\caption[]{(a) Diagrams of the reducuble $G\widetilde{W}$ functional 
$\Phi_{c}^{(r)}$. Full lines are used for $G$, zig-zag lines are used for $D$, empty 
circles are used for $g$, wavy lines are used for $v$ and gluon lines 
are used for $\widetilde{v}$. (b) Electronic self-energy in terms of the 
doubly screened interaction $\widetilde{W}$. (c) Bosonic self-energy in 
terms of the doubly screened response function $\chi$.}
\label{Fig1}
\end{figure}

It is fundamental to observe that the GKBA GFs in Eqs.~(\ref{eq:e:gkba},\ref{eq:b:gkba}) 
are mean-field like GFs. The theory can therefore be  
improved in a conserving fashion by calculating $\cG^{e}$ and $\cG^{b}$ from the  {\em 
reducible} Baym functional $\Phi_{c}^{(r)}$~\cite{stefanucci_nonequilibrium_2013}. Let $\Phi_{c}^{(r)}$ be the $G\widetilde{W}$ 
functional in Fig.~\ref{Fig1}(a) where $\widetilde{\mv}=\mv+ \bcallg^{\dag}\mD
\bcallg$. This functional is reducible with respect to $\mD$ but 
no double counting occurs 
 if $\mD$ is evaluated from Eq.~(\ref{eq:b:gkba}).
Remarkably, a time-linear GKBA+ODE scheme can be formulated
in this case too.  The zeroth order 
contribution (in $g$) is the well known $GW$ approximation
while the second-order contribution corresponds 
to the aforementioned approximation with dynamically screened 
$\bcallg^{d}$, henceforth  $G\widetilde{W}^{(2)}$.

% Let $\Phi_{c}$ be the $G\tilde{W}$ functional in Fig.~\ref{Fig1}(a). The zero-th order 
% contribution (in $g$) is the standard $GW$ approximation
% while the second-order contribution corresponds 
% to the aforementioned approximation with dynamically screened 
% $\bcallg^{d}$, henceforth  $G\tilde{W}^{(2)}$.
The high-order GFs of the doubly screened $G\widetilde{W}$ scheme
follow from Eqs.~(\ref{hogs}) with $\Phi_{c}^{(r)}$ in place of 
$\Phi_{c}$
(time integrals are over the Keldysh contour)
\begin{subequations}
\begin{align}
  \mcG^{e}(t)&=-i\int d\bar{t}d\bar{t}'\mchi(t,\bar{t})
  \widetilde{\mv}(\bar{t},\bar{t}')\mchi^{0}(\bar{t}',t^{+}),
  \\
  \mcG^{b}(t)&=i\int d\bar{t} \mD(t,\bar{t})\bcallg(\bar{t})
  \mchi(\bar{t},t^{+}).
\end{align}
\label{dsgs}%
\end{subequations}
In analogy with $\mchi$ and $\mv$ we have defined 
$\mcG^{e}$ as a matrix in  the two-electron space  with elements 
$\cG^{e}_{\mu\nu}=\cG^{e}_{\begin{subarray}{c}mj\\ 
ni\end{subarray}}=\mathcal{G}^{e}_{imjn}$, and  
in analogy with $\bcallg$ we have defined $\mcG^{b}$ as a 
matrix with elements 
$\cG^{b}_{\mu\nu}=\cG^{b}_{\mu,\begin{subarray}{c}j\\i\end{subarray}}=\cG^{b}_{\mu,ij}$. 
The solution of the EOM  
(\ref{1PEOM}) with  $\mcG^{e}$ and $\mcG^{b}$ from 
Eqs.~(\ref{dsgs}) is equivalent to solving
the KBE with electronic (nonskeletonic) self-energy $\Sigma^{e}=-iG\widetilde{W}$, see 
Fig.~\ref{Fig1}(b), and 
bosonic (reducible) self-energy $\mSgm^{b}=\bcallg\mchi\bcallg^{\dagger}$, see Fig.~\ref{Fig1}(c).
The nonskeletonicity and reducibility is equivalent to dressing of the GKBA $\mD$.

% The underlying diagrammatic approximation corresponds to
% the second-order 
% expansion in $g$ of a $GW$-like self-energy where  
% the bare $e$-$e$ interaction $\mv$  is 
% renormalized by processes involving the emission/absorption of  
% virtual bosons,  i.e., 
% $\mv\to\tilde{\mv}=\mv+ \bcallg^{\dag}\mD
% \bcallg$ and hence $\mW\to 
% \tilde{\mW}=\tilde{\mv}(1+\mchi^{0}\tilde{\mW})$. 
% Remarkably, a time-linear GKBA+ODE scheme can be formulated
% in this case too. 

The GKBA in Eqs.~(\ref{eq:e:gkba},\ref{eq:b:gkba})
can be used to transform $\mcG^{e}$ and $\mcG^{b}$ into functionals 
of $\rho^{<}$ and $\bgg^<$, see Appendix~\ref{appgkbagegb}, thus closing the EOM for these 
quantities. Interestingly, however, the EOM for these high-order 
GFs form a closed system. We separate 
the two-particle GF into a purely electronic part 
$\mcG^{ee}\equiv \mcG^{e}|_{g=0}$ (diagrams with no $e$-$b$ vertices)
and a rest $\mcG^{eb}$, hence 
$\mcG^{e}=\mcG^{ee}+\mcG^{eb}$, and show in
Appendix~\ref{appEOMgegb} that (omitting the dependence on the time 
variable) 
\begin{subequations}
\begin{align}
i\frac{d}{dt}\mcG^{ee}&=-\mPsi^{e}+\mh^{e}_{\rm 
eff}\mcG^{ee}-\mcG^{ee}\mh^{e\dagger}_{\rm eff},
\label{EOMGee}
\\
i\frac{d}{dt}\mcG^{eb}&=\mrho^\Delta\bcallg^{\dagger}\mcG^{b}
-\mcG^{b\dagger}\bcallg\mrho^\Delta+\mh^{e}_{\rm 
eff}\mcG^{eb}-\mcG^{eb}\mh^{e\dagger}_{\rm eff},
\label{EOMGeb}
\\
i\frac{d}{dt}\mcG^{b}&=-\mPsi^{b}-\bga\bcallg\mcG^{e}-
\mcA\bcallg\mrho^\Delta+\mh^{b}\mcG^{b}-\mcG^{b}\mh^{e\dagger}_{\rm 
eff},
\label{EOMGb}
\\
i\frac{d}{dt}\mcA&=\mcG^{b}\bcallg^{\dagger}\bga-\bga\bcallg\mcG^{b\dagger}+
\mh^{b}\mcA-\mcA\mh^{b},
\label{EOMA}
\end{align}
\label{EOMG}%
\end{subequations}
where $\mcA$ is an auxiliary quantity needed to close the EOM.
The driving terms $\mPsi^{e}$ and $\mPsi^{b}$  are 
functionals of $\rho^{<}$ and $\bgg^<$. They have been already 
encountered in 
Refs.~\cite{schlunzen_achieving_2020,karlsson_fast_2021} in the 
context of the simpler $GW$ and $GD$ approximations. In particular
\begin{align}
  \mPsi^{e}(t)&\equiv \mrho^{>}(t)\mv(t)\mrho^{<}(t)-\mrho^{<}(t)\mv(t) \mrho^{>}(t).
  \label{psie}
  \\
   \mPsi^{b}(t)&\equiv 
   \bgg^{>}(t)\bcallg(t)\mrho^{<}(t)-\bgg^{<}(t)\bcallg(t) \mrho^{>}(t),
  \label{Psi:b:def}
\end{align} 
and $\mh^{e}_{\rm eff}=\mh^{e}-\mrho^\Delta\mv$ with 
$\mrho^\Delta=\mrho^{>}-\mrho^{<}$. The matrices $\mh^{e}$ and 
$\mrho^{\gtrless}$ in the two-electron space (hence represented by 
boldface letters) are defined with 
elements $h^{e}_{\mu\nu}=h^{e}_{\begin{subarray}{c}ij\\ mn\end{subarray}}
=h^{e}_{ij}\delta_{nm}-\delta_{ij}h^{e}_{nm}$ and 
$\rho^{\lessgtr}_{\mu\nu}=\rho^{\lessgtr}_{\begin{subarray}{c}ij\\ mn\end{subarray}}= 
\rho^{\lessgtr}_{ij}\rho^{\gtrless}_{nm}$.

Equations~(\ref{1PEOM},\ref{EOMG}) together with the Ehrenfest 
equation for $\phi_{\mu}$, see below, form a system of seven first-order ODE that 
can be conveniently solved numerically using a time-stepping 
algorithm. This is the first main result of our work. The 
$G\widetilde{W}^{(2)}$
approximation is easily derived by discarding terms of 
order higher than $g^{2}$. In Appendix~\ref{appgkbagegb} we show that 
$\mcG^{eb}={\mathcal O}(g^{2})$, $\mcG^{b}={\mathcal O}(g)$ and 
$\mcA={\mathcal O}(g^{2})$. Hence to second order in $g$ the r.h.s. 
of Eq.~(\ref{EOMGb}) can be calculated with $\bcallg\mcG^{e}\to\bcallg\mcG^{ee}$ 
and $\bcallg\mcA\to 0$; this implies that in $G\widetilde{W}^{(2)}$
the EOM for $\mcA$ decouples. We also observe that the EOM 
in the $GD$ approximation, see Ref.~\cite{karlsson_fast_2021}, 
are recovered from the $G\widetilde{W}^{(2)}$
 method upon setting $v=0$ (in this case we are left   with only the 
equation for $\mcG^{b}$). The EOM in the $GW$ 
approximation~\cite{schlunzen_achieving_2020,joost_g1-g2_2020,pavlyukh_photoinduced_2021} are instead recovered 
from the full $G\widetilde{W}$ method upon setting $g=0$  (in this 
case we are left with only the 
equation for $\mcG^{ee}$).

\begin{figure*}[t]
\centering\includegraphics[width=18cm]{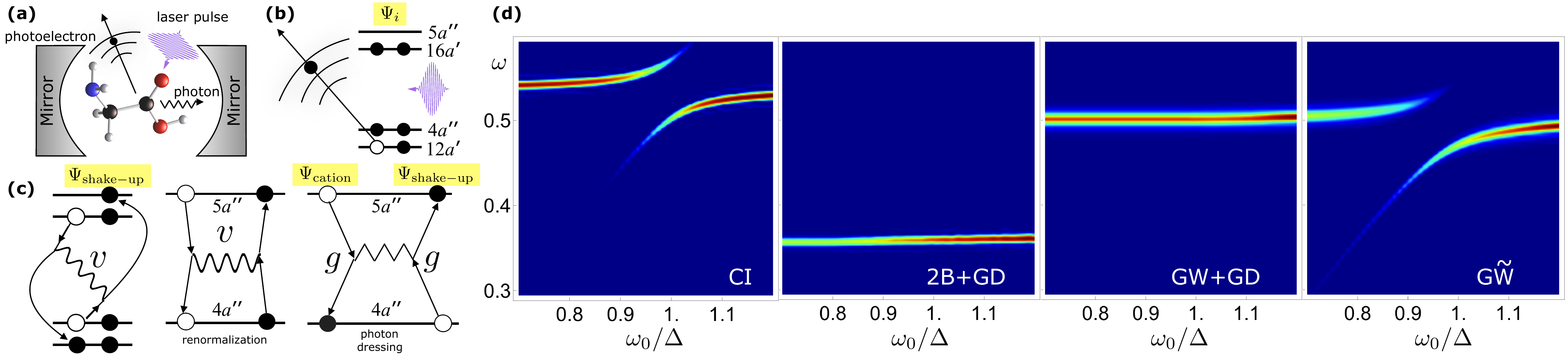}
\caption[]{(a) Illustration of the gedanken experiment. A Gly molecule is 
ionized by a laser pulse and a cavity-photon is emitted. (b) The four 
MOs involved in the charge migration  of Gly when the electron is 
ionized from the $12a'$ MO. Electrons (black dots) on the MOs 
identify the state $\Psi_{i}$ after ionization. (c) Shake-up process 
leading to state $\Psi_{\rm shake-up}$ (left); scattering between 
electrons in the $4a''$ and $5a''$ MOs responsible for a sizable 
renormalization of the energy of the shake-up state (middle); electron-photon 
scattering leading to transition $\Psi_{\rm shake-up}\leftrightarrow \Psi_{\rm cation}$ (right).  (d) Spectrograms 
of the occupancy of the $12a'$ MO in different schemes.}
\label{gedanken}
\end{figure*}

{\em Combining different methods:} The treatment of pure electronic 
correlations is not limited to the $GW$ approximation. By properly 
modifying the index order of the matrices $\mcG^{ee}$, 
$\mrho^{\lessgtr}$, $\mh^{e}$ and $\mv$ in Eq.~(\ref{EOMGee}) we can explore a large 
variety of methods~\cite{pavlyukh_photoinduced_2021}. They include the one-bubble or second-order 
direct (2B$^{d}$), second-order exchange (2B$^{x}$), $GW$, exchange-only 
$GW$ ($XGW$), $GW$ plus exchange ($GW+X$), $T$-matrix in the 
particle-hole channel ($T^{ph}$),  exchange-only $T^{ph}$ 
($XT^{ph}$), $T^{ph}$ plus exchange ($T^{ph}+X$), $T$-matrix in the 
particle-particle channel ($T^{pp}$) and exchange-only $T^{pp}$ 
($XT^{pp}$), 
see Appendix~\ref{appec}.
Let ``$c$'' be the index for  one of these correlated methods
and let us 
denote by $\cG^{ee(c)}_{imjn}$ the corresponding two-particle GF. 
%
% Different methods can be combined by paying attention to double 
% coutings. We can therefore include several
% type of correlation effects {\em simultaneously} if the 
% two-particle Green's function $\cG^{ee}$ is evaluated according to
Different methods can be combined to {\em simultaneously} include several types of 
correlation effects if the 
two-particle GF $\cG^{ee}$ is evaluated according to
\begin{align}
\cG^{ee}_{imjn}(t)=\sum\nolimits_{c}n_{c}\cG^{ee(c)}_{imjn}(t\,).
\label{geesumcgeec}
\end{align}
In Appendix~\ref{appec} we discuss how to choose the integers $n_{c}$
to avoid double countings.
Decorating the electronic two-particle matrices $\mrho^{\lessgtr}$, $\mh^{e}$ and $\mv$ 
in the EOM for $\mcG^{ee(c)}$ with the superscript~$c$, 
the whole GKBA+ODE toolbox for interacting electrons and bosons 
can then be summarized as
(omitting the dependence on 
the time variable)
\begin{widetext}
\begin{subequations}
\begin{align}
i\frac{d}{dt}\phi_\mu&=  h^{b}_{\mu\nu}\phi_\nu 
  +\sum\nolimits_{\nu,ij}\alpha_{\mu\nu}g_{\nu,ij} \rho_{ji},
  \label{GKBA-ODE1}
\\
i\frac{d}{dt}\rho^<_{lj}&=
\Bigl\{\sum_{i}h^{e}_{li}\rho^<_{ij}
-\sum_{imn} v_{lnmi}
  \big[\cG^{ee}_{imjn}+s_{1}d\cG^{eb}_{imjn}\big]
  +d \sum_{\mu,i} 
  g_{\mu,li}\,\cG^{b}_{\mu,ij}\Bigr\}-\{l\leftrightarrow j\}^{\ast},
\label{GKBA-ODE2}
\\
i\frac{d}{dt}\gamma_{\mu\nu}^<&= 
\Bigl\{\sum_{\beta}
h^{b}_{\mu\beta}\gamma^<_{\beta\nu} 
  +d\sum_{\beta,mn} \alpha_{\mu\beta}
  g_{\beta,mn}\,\cG^{b}_{\nu,nm}\Bigr\}-\{\mu\leftrightarrow\nu\}^{\ast},
  \label{GKBA-ODE3}
\\  
%\end{align}
%\begin{align}
i\frac{d}{dt}\mcG^{ee(c)}&=-\mPsi^{e(c)}+\mh^{e(c)}_{\rm 
eff}\mcG^{ee(c)}-\mcG^{ee(c)}\mh^{e(c)\dagger}_{\rm eff},
\label{GKBA-ODE4}
\\
i\frac{d}{dt}\mcG^{eb}&=\mrho^{\Delta(GW)}\bcallg^{\dagger}\mcG^{b}
-\mcG^{b\dagger}\bcallg\mrho^{\Delta(GW)}+\mh^{e(GW)}_{\rm 
eff}\mcG^{eb}-\mcG^{eb}\mh^{e(GW)\dagger}_{\rm eff},
\label{GKBA-ODE5}
\\
i\frac{d}{dt}\mcG^{b}&=-\mPsi^{b}-s_{1}\bga\bcallg\big[\mcG^{ee(GW)}+s_{2}\mcG^{eb}\big]-
s_{1}s_{2}\mcA\bcallg\mrho^{\Delta(GW)}+\mh^{b}\mcG^{b}-\mcG^{b}
\big[\mh^{e(GW)}-s_{1}\mrho^{\Delta(GW)}\mv^{(GW)}\big],
\label{GKBA-ODE6}
\\
%\end{align}
%\begin{align}
i\frac{d}{dt}\mcA&=\mcG^{b}\bcallg^{\dagger}\bga-\bga\bcallg\mcG^{b\dagger}+
\mh^{b}\mcA-\mcA\mh^{b}.
\end{align}
\label{GKBA-ODE}%
\end{subequations}
\end{widetext}
The control parameters $d$, $s_{1}$ and $s_{2}$ refer to the 
treatment of $e$-$b$ correlations. 
The Ehrenfest approximation is recovered for $d=0$ -- in this case 
the only equations to solve are those for the 
displacements and momenta, i.e., Eq.~(\ref{GKBA-ODE1}), and the 
electronic equations (\ref{GKBA-ODE2}) and (\ref{GKBA-ODE4}).
$e$-$b$ correlations are included choosing $d=1$. In this case 
we 
can set $(s_{1},s_{2})=(0,0)$ ($GD$), 
$(s_{1},s_{2})=(1,0)$ ($G\widetilde{W}^{(2)}$) and 
$(s_{1},s_{2})=(1,1)$ ($G\widetilde{W}$). The number of 
equations (\ref{GKBA-ODE4}) 
depends on the chosen treatment of
electronic correlations, i.e., on the values of $n_{c}$'s. If $n_{c}=0$ the 
corresponding $\mcG^{ee(c)}$ is not needed. The only exception is for 
$c=GW$:  if $s_{1}=1$ then the EOM for 
$\mcG^{ee(GW)}$ must be solved even for $n_{GW}=0$, see Eq.~(\ref{GKBA-ODE6}).
The GKBA+ODE toolbox in Eqs.~(\ref{GKBA-ODE}) generalizes the one 
published in Ref.~\cite{pavlyukh_time-linear_2021} in two ways (i) it 
includes the $G\widetilde{W}^{(2)}$ and $G\widetilde{W}$ methods and (2) it 
allows for combining 
different treatments of electronic correlations, for a total of  
$2^{12}$ distinct diagrammatic methods, see Appendix~\ref{appmerging}. 
This is the second 
main result of our work.

{\em Charge migration in a cavity:} We consider the Gly I conformer of the glycine 
molecule and study the correlation-induced charge migration due to the removal of an electron from 
the 12$a'$ molecular orbital (MO), see Fig.~\ref{gedanken}(b). In free space this case has 
been investigated at 
length~\cite{kuleff_multielectron_2005,kuleff_charge_2007,cooper_single-photon_2013,perfetto_first-principles_2019,pavlyukh_photoinduced_2021}.  
Coulomb interaction is responsible for a {\em shake-up} process where an 
electron from the $16a'$ MO fills the photo-hole and another electron is 
promoted from the $4a''$ MO to the initially empty $5a''$ MO, 
left of Fig.~\ref{gedanken}(c). 
We refer to our previous works for the electronic structure and
basis 
representation~\cite{perfetto_first-principles_2019,perfetto_cheers:_2018}. 
In Ref.~\cite{pavlyukh_photoinduced_2021} we 
showed that the energy of the shake-up state is 
strongly {\em renormalized} 
by the exchange interaction between electrons in the $4a''$ and $5a''$ MOs, middle 
of Fig.~\ref{gedanken}(c), and that capturing this renormalization 
requires a $GW$ treatment. Here we analyze how the dynamics is affected by 
a single cavity-mode that couples the shake-up state $\Psi_{\rm 
shake-up}$ to the 
lowest-energy cationic state $\Psi_{\rm cation}$ (one hole in $16a'$ MO), right of 
Fig.~\ref{gedanken}(c).

Let $\Delta=E_{\rm shake-up}-E_{i}=0.522$~a.u. be the energy difference between 
$\Psi_{\rm shake-up}$  and the state $\Psi_{i}$ 
of Gly just after photo-ionization. In 
Fig.~\ref{gedanken}(d) we show the Fourier transform of the occupancy 
of the $12a'$ MO for different frequencies $\omega_{0}$ of the cavity mode. 
The coupling $g=\lambda d_{4a'',5a''}\sqrt{\omega_{0}}$ is proportional to 
the dipole moment $d_{4a'',5a''}$ between the MOs involved in the 
transition  $\Psi_{\rm shake-up}\to\Psi_{\rm cation}$.  The 
electron-photon coupling strength $\lambda$ is determined by the mode 
wavefunction at the location of the 
molecule~\cite{yang_quantum-electrodynamical_2021}. We take 
$d_{4a'',5a''}=0.125$~a.u. as the average dipole 
moment along three orthogonal direction and choose 
$\lambda=0.212$~a.u.. Details on the numerical simulations can be found 
in Appendix~\ref{appnum}.

The first panel of Fig.\ref{gedanken}(d) displayes the Configuration 
Interaction (CI) spectrogram. 
For $\omega_{0}\ll \Delta$ cavity-photons are hardly emitted and the only 
possible transition is $\Psi_{i}\leftrightarrow \Psi_{\rm shake-up}$.
Correspondingly, the spectrum  has only one peak 
at frequency $\Delta_{\rm CI}=0.544~{\rm a.u.}~\simeq \Delta$. As $\omega_{0}$ approaches $\Delta$ an 
Autler-Townes doublet of entangled electron-photon many-body states
becomes visible~\cite{autler_stark_1955,perfetto_exact_2015}.
It is due to the {\em photon-dressing} of 
the cationic state which makes the transition 
$\Psi_{i}\leftrightarrow \Psi_{\rm cation}$ bright and dominant when 
$\omega_{0}>\Delta$. 

For a diagrammatic approximation to reproduce CI, the electronic self-energy must 
account for all three mechanisms illustrated in Fig.~\ref{gedanken}(c).
In the second panel of Fig.\ref{gedanken}(d) we 
report the 2B+$GD$ spectrogram. This approximation captures only 
the shake-up process, thereby yielding a 
$\omega_{0}$-independent structure at energy $\Delta_{\rm 2B}=0.356$~a.u..
As expected~\cite{pavlyukh_photoinduced_2021}, the $GW+GD$ method 
renormalizes $\Delta_{\rm 2B}$ to $\Delta_{GW}=0.503\simeq \Delta_{\rm 
2B}+2v^{x}_{4a'',5a''}$, 
see third panel, where $v^{x}_{4a'',5a''}=0.08$~a.u. is the exchange Coulomb 
integral responsible for the scattering in Fig.~\ref{gedanken}(c) 
(middle). Achieving the CI value $\Delta$
calls for vertex corrections which, however, are beyond the current 
GKBA+ODE formulation. The most severe deficiency 
of the $GW+GD$ spectrogram is the absence of the Autler-Townes doublet.
In fact, photon-dressing requires a non-perturbative 
 treatment in the $e$-$b$ coupling like the $G\widetilde{W}$ 
method. The $G\widetilde{W}$ spectrogram is shown in the fourth panel. 
Although the intensity of the low-$\omega_{0}$ peak is weaker than 
in CI, the improvement over $GW+GD$ is quantitatively and 
qualitatively substantial. 

In conclusion, we have extended the time-linear GKBA+ODE formulation for 
interacting fermions and bosons to 
the doubly screened $G\widetilde{W}$ method, and shown how to combine different diagrammatic 
approximations to account for multiple correlation effects 
simultaneously while preserving all conserving properties. 
The case of correlation-induced charge migration of glycine in an optical cavity 
exemplifies the superiority of $G\widetilde{W}$ over current state-of-the-art 
approaches.
We emphasize that the scaling of a $G\widetilde{W}$ calculation with the system size is the 
same as for $GW$, thus making the method potentially available for real-time 
first-principles simulations of 
finite~\cite{perfetto_cheers:_2018,pavlyukh_photoinduced_2021} and 
extended~\cite{sangalli_many-body_2019,perfetto_real-time_2022} systems.
Last but not least the GKBA+ODE formulation lends itself to 
studies of multiscale phenomena  through the 
implementation of adaptive time-stepping algorithms.

\begin{acknowledgments}
  We acknowledge the financial support from MIUR PRIN (Grant No. 20173B72NB), from INFN
  through the TIME2QUEST project, and from Tor Vergata University through the Beyond
  Borders Project ULEXIEX. We also acknowledge useful discussions 
  with Andrea Marini.
\end{acknowledgments}

\appendix

\section{GKBA form of $\cG^{e}$ and $\cG^{b}$}
\label{appgkbagegb}

We here work out the GKBA expression for the high-order GFs 
in Eq.~(\ref{Gedef}) and (\ref{Gbdef}). Let us start from 
$\mcG^{e}$. Using the Langreth rules we find
\begin{align}
\mcG^{e}(t)&=\!\!\int\!\! d\bar{t}d\bar{t}'\big[
\mchi^{R}(t,\bar{t})\tilde{\mv}^{R}(\bar{t},\bar{t}')\mchi^{0,<}(\bar{t}',t)
\nonumber\\
&+\!\mchi^{R}(t,\bar{t})\tilde{\mv}^{<}(\bar{t},\bar{t}')\mchi^{0,A}(\bar{t}',t)
+\!\mchi^{<}(t,\bar{t})\tilde{\mv}^{A}(\bar{t},\bar{t}')\mchi^{0,A}(\bar{t}',t)
\big],
\label{Ge1}
\end{align}
where all intergrals are now over the real axis.
From the RPA equation 
$\mchi=\mchi^{0}+\mchi\tilde{\mv}\mchi^{0}$ (on the Keldysh contour) we 
can easily extract the retarded ($R$), advanced ($A$), lesser ($<$) and 
greater ($>$) components
\begin{subequations}
\begin{align}
\mchi^{R}&=\mchi^{0,R}+\mchi^{0,R} \cdot 
\tilde{\mv}^{R}\cdot\mchi^{R},
\label{RPAR}
\\
\mchi^{A}&=\mchi^{0,A}+\mchi^{0,A} \cdot 
\tilde{\mv}^{A}\cdot\mchi^{A},
\label{RPAA}
\\
\mchi^{\lessgtr}&=(\bgd+\mchi^{R}\cdot\tilde{\mv}^{R})\cdot
\mchi^{0,\lessgtr}\cdot(\bgd+\tilde{\mv}^{A}\cdot\mchi^{A} )
+\mchi^{R}\cdot\tilde{\mv}^{\lessgtr}\cdot\mchi^{A},
\label{RPA<>}
\end{align}
\end{subequations}
where the ``$\cdot$'' symbol signifies a convolution on the real axis.
The explicit expression for the components of the renormalized 
interaction is
\begin{subequations}
\begin{align}
\tilde{\mv}^{R}(t,t')&=\mv(t)\delta(t,t')+\bcallg^{\dagger}(t)\mD^{R}(t,t')\bcallg(t'),
\\
\tilde{\mv}^{A}(t,t')&=\mv(t)\delta(t,t')+\bcallg^{\dagger}(t)\mD^{A}(t,t')\bcallg(t'),
\\
\tilde{\mv}^{\lessgtr}(t,t')&=\bcallg^{\dagger}(t)\mD^{\lessgtr}(t,t')\bcallg(t').
\end{align}
\end{subequations}
In Ref.~\cite{pavlyukh_time-linear_2021} we have shown that the GKBA 
form of $\mchi^{0,R/A}$ and $\mchi^{0,\lessgtr}$ is
\begin{subequations}
\begin{align}
\mchi^{0,R}(t,t')&=\mP^{R}(t,t')\mrho^{\Delta}(t'),
\label{chi0r}
\\
\mchi^{0, A}(t,t')&=\mrho^{\Delta}(t)\mP^{A}(t,t'),
\label{chi0a}
\\
\mchi^{0, \lessgtr}(t,t')&=\mP^{R}(t,t')\mrho^{\lessgtr}(t')-
\mrho^{\lessgtr}(t)\mP^{A}(t,t'),
\label{chi0><}
\end{align}
\label{chi0ra><}%
\end{subequations}
where the bare propagator 
$\mP^{R}(t,t')=[\mP^{A}(t',t)]^{\dag}$ fulfills the EOM
\begin{align}
  i\frac{d}{dt}\mP^R(t,t')&=\mh^{e}(t)\mP^R(t,t'),
\label{eq:EOM:P}
\end{align}
with boundary condition 
\begin{align}
\mP^R(t^{+},t)=i\mathbb{1}
\label{bcPR}
\end{align}
and $\mP^R(t,t')=0$ for $t<t'$.
Substituting Eqs.~(\ref{chi0r}) and (\ref{chi0a}) into Eqs.~(\ref{RPAR}) and 
(\ref{RPAA}) we find
\begin{subequations}
\begin{align}
\mchi^{R}(t,t')&=\mPi^{R}(t,t')\mrho^{\Delta}(t'),
\\
\mchi^{A}(t,t')&=\mrho^{\Delta}(t)\mPi^{A}(t,t'),
\label{chiA}
\end{align}
\label{chira}%
\end{subequations}
where the dressed propagator $\mPi^{R}(t,t')=\left[\mPi^{A}(t',t)\right]^\dagger$ fulfills
the RPA equation
\begin{subequations}
\begin{align}
  \mPi^{R}-\mP^{R}&=\mPi^{R}\cdot\mrho^\Delta\tilde{\mv}^{R}\cdot\mP^{R}
  \nonumber \\
  &=\mP^{R}\cdot \mrho^\Delta\tilde{\mv}^{R}\cdot\mPi^{R},\\
  \mPi^{A}-\mP^{A}&=\mPi^{A}\cdot\tilde{\mv}^{A}\mrho^\Delta\cdot \mP^{A}
 \nonumber \\
 &=\mP^{A}\cdot \tilde{\mv}^{A}\mrho^\Delta\cdot\mPi^{A}.
\end{align}
\label{eq:rpa:PiR}
\end{subequations}
For later purposes we find convenient to define the purely electronic 
dressed propagator
\begin{subequations}
\begin{align}
  \mPi^{eR}-\mP^{R}&=\mPi^{eR}\mrho^\Delta\mv\cdot\mP^{R}
  \nonumber \\
  &=\mP^{R} \cdot\mrho^\Delta\mv\mPi^{R},\\
  \mPi^{eA}-\mP^{A}&=\mPi^{eA}\cdot\mv\mrho^\Delta \mP^{A}
  \nonumber \\
  &=\mP^{A}\cdot \mv\mrho^\Delta\mPi^{eA},
\end{align}
\label{eq:rpa:PieR}
\end{subequations}
in terms of which Eqs.~(\ref{eq:rpa:PiR}) can be rewritten as
\begin{subequations}
\begin{align}
  \mPi^{R}-\mPi^{eR}&=\mPi^{R}\cdot\mrho^\Delta\bcallg^{\dagger}
  \mD^{R}\bcallg\cdot\mPi^{eR}
  \nonumber \\
  &=\mPi^{eR}\cdot\mrho^\Delta\bcallg^{\dagger}
  \mD^{R}\bcallg\cdot\mPi^{R},
  \\
  \mPi^{A}-\mPi^{eA}&=\mPi^{A}\cdot\bcallg^{\dagger}
  \mD^{A}\bcallg\mrho^\Delta\cdot \mPi^{eA}
  \nonumber \\
  &=\mPi^{eA}\cdot \bcallg^{\dagger}
  \mD^{A}\bcallg\mrho^\Delta\cdot\mPi^{A}.
\end{align}
\label{eq:rpa:PiR2}
\end{subequations}
Substituting  Eqs.~(\ref{chira}) and Eq.~(\ref{chi0><}) into 
Eq.~(\ref{RPA<>}) we find
\begin{widetext}
\begin{align}
  \mchi^{\lessgtr}
 =\mPi^R\mrho^{\lessgtr}\cdot\big(\bgd+\tilde{\mv}^{A}\mrho^\Delta\cdot \mPi^A\big)
-\big(\bgd+\mPi^R\cdot\mrho^\Delta\tilde{\mv}^{R}\big)\cdot\mrho^{\lessgtr}\mPi^A
 +\mPi^R\cdot\mrho^\Delta\tilde{\mv}^{\lessgtr}\mrho^{\lessgtr}\cdot\mPi^A.
  \label{eq:chi:gtrless}
\end{align}

The GKBA form of the response function, i.e., Eqs.~(\ref{chi0ra><}), (\ref{chira}) and 
(\ref{eq:chi:gtrless}) can now be transferred in 
Eq.~(\ref{Ge1}). After some algebra we obtain
\begin{align}
\mcG^{e}(t)=\!\int\!\! d\bar{t}d\bar{t}'\;\mPi^R(t,\bar{t})
\mPsi(\bar{t},\bar{t}')\mPi^A(\bar{t}',t),
\label{Ge2}
\end{align}
where
\begin{align}
\mPsi(t,t')=\mPsi^{e}(t)\delta(t,t')+   
\mrho^\Delta(t)\bcallg^{\dagger}(t)
\mD^{R}(t,t')\bcallg(t')\mrho^{<}(t')
-\mrho^<(t)\bcallg^{\dagger}(t)
\mD^{A}(t,t')\bcallg(t')\mrho^{\Delta}(t')
-\mrho^\Delta(t)\bcallg^{\dagger}(t)
\mD^{<}(t,t')\bcallg(t')\mrho^{\Delta}(t').
\end{align}
In this equation it appears the driving term $\mPsi^{e}$ defined in 
Eq.~(\ref{psie}). Using the GKBA for bosons in Eq.~(\ref{eq:b:gkba}) 
and taking into account that $\bgg^{>}-\bgg^{<}=\bga$ and that 
$\bga^{2}=1$ we can rewrite $\mPsi$ as
\begin{align}
\mPsi(t,t')&=\mPsi^{e}(t)\delta(t,t')+\mrho^\Delta(t)\bcallg^{\dagger}(t)
\mD^{R}(t,t')\bga\mPsi^{b}(t')-\mPsi^{b\dagger}(t)\bga\mD^{A}(t,t')\bcallg(t')
\mrho^\Delta(t'),
\label{Psi2}
\end{align}
where $\mPsi^{b}$ is the driving term defined in 
Eq.~(\ref{Psi:b:def}). Inserting Eq.~(\ref{Psi2}) into 
Eq.~(\ref{Ge2}) and taking into account Eqs.~(\ref{eq:rpa:PiR2}) to 
isolate the purely electronic part $\mcG^{ee}=\mcG^{e}|_{g=0}$ which 
does not contain explicitly $e$-$b$ vertices we 
obtain
\begin{align}
\mcG^{ee}(t)=i\big[\mPi^{eR}\cdot\mPsi^{e}\mPi^{eA}\big](t,t)
\label{Gee1}
\end{align}
and
\begin{align}
\mcG^{eb}(t)=i\big\{\mPi^{eR}\cdot\big[&\mrho^\Delta\bcallg^{\dagger}
  \mD^{R}\bcallg\cdot\mPi^{R}\mPsi^{e}+
  \mPsi^{e}\mPi^{A}\cdot\bcallg^{\dagger}
  \mD^{A}\bcallg\mrho^\Delta 
  +\mrho^\Delta\bcallg^{\dagger}
  \mD^{R}\bcallg\cdot\mPi^{R}\cdot\mPsi^{e}
  \mPi^{A}\cdot\bcallg^{\dagger}
  \mD^{A}\bcallg\mrho^\Delta 
  \nonumber \\
  +&(\bgd+\mrho^\Delta\bcallg^{\dagger}\mD^{R}\bcallg\cdot\mPi^{R})\cdot
  (\mrho^\Delta\bcallg^{\dagger}
  \mD^{R}\bga\mPsi^{b}-\mPsi^{b\dagger}\bga\mD^{A}\bcallg\mrho^\Delta)\cdot
  (\bgd+\mPi^{A}\cdot\bcallg^{\dagger}
  \mD^{A}\bcallg\mrho^\Delta) 
  \big]\cdot\mPi^{eA}\big\}(t,t)
  \label{Geb1}
\end{align}
Notice that $\mcG^{ee}={\mathcal O}(g^{0})$ whereas 
$\mcG^{eb}={\mathcal O}(g^{2})$.

Let us now come to $\mcG^{b}$ in Eq.~(\ref{Gbdef}). The 
Langreth rules yield
\begin{align}
\mcG^{b}(t)=i\int d\bar{t}\big[
\mD^{<}(t,\bar{t})\bcallg(t)\mchi^{A}(\bar{t},t)+
\mD^{R}(t,\bar{t})\bcallg(t)\mchi^{<}(\bar{t},t)\big].
\end{align}
Using the GKBA form of $\mD^{<}$ [Eq.~(\ref{eq:b:gkba})], 
$\mchi^{A}$ [Eq.~(\ref{chiA})] and $\mchi^{<}$ 
[Eq.~(\ref{eq:chi:gtrless})], after some algebra we find
\begin{align}
\mcG^{b}(t)=-i\big\{\big[\mD^{R}\cdot (\bga\mPsi^{b}+
\bcallg\mPi^{R}\cdot\mPsi)\cdot
  (\bgd+\mPi^{A}\cdot\bcallg^{\dagger}
  \mD^{A}\bcallg\mrho^\Delta) \big]\cdot
  \mPi^{eA}\big\}(t,t),
  \label{Gb1}
\end{align}
and hence $\mcG^{b}={\mathcal O}(g)$.

\end{widetext}

\section{Equations of motion for $\cG^{e}$ and $\cG^{b}$}
\label{appEOMgegb}

Due to the presence of retarded propagators on the left and advanced 
propagators on the right the high-order GFs 
are convolutions of the form
\begin{align}
\cG(t)=\int^{t} \!\!d\bar{t} \!\int^{t}\!\! d\bar{t}'
R(t,\bar{t})K(\bar{t},\bar{t}')A(\bar{t}',t).
\end{align}    
The derivative of $\cG$ with respect to time is given by
\begin{align}
\frac{d\cG(t)}{dt}&=R(t^{+}\!,t)\!\!\int^{t}\!\! \!d\bar{t}
K(t,\bar{t})A(\bar{t},t)+\!\!
\int^{t} \!\!\!d\bar{t} 
R(t,\bar{t})K(\bar{t},t)A(t,t^{+})
\nonumber \\
&+\int^{t} \!\!d\bar{t} \!\int^{t}\!\! d\bar{t}'
\frac{dR(t,\bar{t})}{dt}K(\bar{t},\bar{t}')A(\bar{t}',t)
\nonumber \\
&+\int^{t} \!\!d\bar{t} \!\int^{t}\!\! d\bar{t}'
R(t,\bar{t})K(\bar{t},\bar{t}')\frac{dA(\bar{t}',t)}{dt}.
\label{dcdt}
\end{align} 
The EOM for the high-order correlators can therefore be inferred from the 
EOM of the propagators and from their values at equal time. From Eq.~(\ref{eq:EOM:P}) it is 
straightforward to derive the EOM for the electronic $\mPi^{eR/A}$ 
defined in Eq.~(\ref{eq:rpa:PieR})
\begin{subequations}
\begin{align}
  i\frac{d}{dt}\mPi^{eR}(t,t')&=\mh^{e}_{\rm eff}(t)\mPi^{eR}(t,t'),
  \quad t>t',
\label{eq:EOM:PieR}
 \\
 i\frac{d}{dt}\mPi^{eA}(t',t)&=-\mPi^{eA}(t',t)\mh^{e}_{\rm eff}(t),
   \quad t>t',
\label{eq:EOM:PieA}
\end{align}
\end{subequations}
where $\mh^{e}_{\rm eff}(t)=\mh^{e}(t)-\mrho^{\Delta}(t)\mv(t)$.
The EOM for the retarded bosonic propagator follows from its 
definition 
\begin{align}
\mD^{R}(t,t')=-i\bga \theta(t-t') T\left\{e^{-i \int_{t'}^t d\tau\, 
\mh^{b}(\tau)}\right\},
\label{eq:dr:hf}
\end{align}
and it reads
\begin{align}
i\frac{d}{dt}\mD^R(t,t')=\mh^{b}(t)\mD^R(t,t').
\end{align}
The equal-time values of $\mPi^{eR/A}$ is the same as the equal-time 
value of $\mP^{R/A}$, i.e., 
$\mPi^{eR}(t^{+},t)=[\mPi^{eA}(t,t^{+})]=i\mathbb{1}$, see 
Eq.~(\ref{bcPR}). The equal-time value of the bosonic propagator is 
instead $\mD^{R}(t^{+},t)=-i\bga$, see Eq.~(\ref{eq:dr:hf}).

Using the relation in Eq.~(\ref{dcdt}) for $\mcG^{ee}$ and 
$\mcG^{eb}$ in 
Eqs.~(\ref{Gee1}) and (\ref{Geb1}) we easily find Eqs.~(\ref{EOMGee}) 
and (\ref{EOMGeb}). The time derivative of $\mcG^{b}$ in 
Eq.~(\ref{Gb1}) yields Eq.~(\ref{EOMGb}) where 
\begin{align}
\mcA(t)&=i\big\{\mD^{R}\cdot\big[\bcallg\mPi^{R}\cdot\mPsi^{e}
\mPi^{A}\bcallg^{\dagger}
\nonumber \\
&+(\bgd+ \bcallg\mPi^{R}\cdot\mrho^\Delta \bcallg^{\dagger}\mD^{R})
\bga\mPsi^{b}\cdot\mPi^{A}\bcallg^{\dagger}
\nonumber \\
&-\bcallg\mPi^{R}\cdot\mPsi^{b\dagger}\bga
 (\bgd+\mD^{A}\bcallg\mrho^\Delta\cdot\mPi^{R}\bcallg^{\dagger})
 \big]\cdot\mD^{A}\big\}(t,t).
\end{align}
Since $\mPsi^{b}={\mathcal O}(g)$ we see that $\mcA={\mathcal 
O}(g^{2})$. The time derivative of $\mcA$ can be easily worked out 
using again the relation  in Eq.~(\ref{dcdt}), and it leads to 
Eq.~(\ref{EOMA}).

\section{Electronic correlated methods}
\label{appec}

\begin{figure}[tbp]
\centering  \includegraphics[width=0.99\columnwidth]{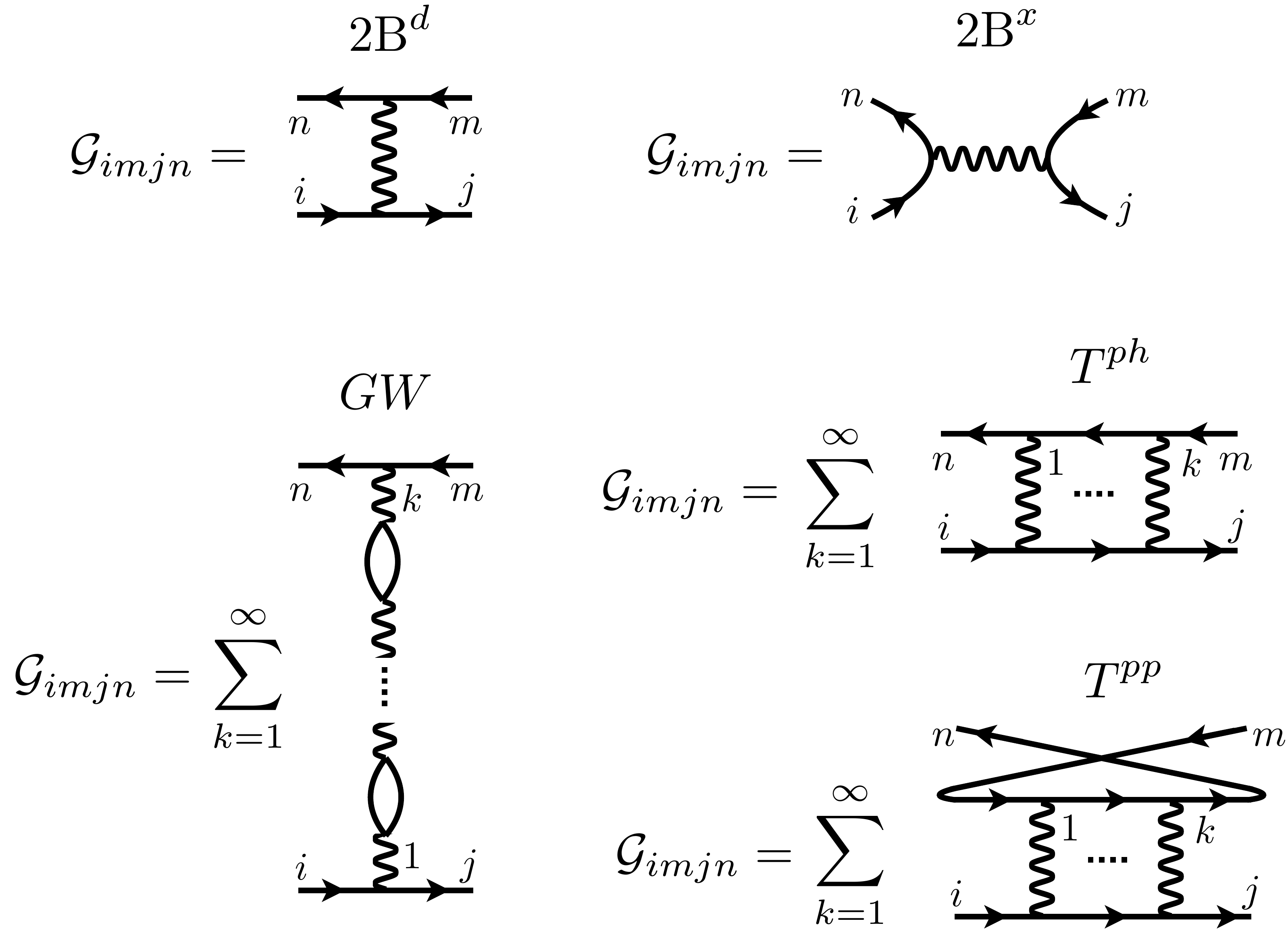}
\caption[]{Top: Diagrams for the 2B$^{d}$ (left) and 2B$^{x}$ 
(right) methods. Bottom: Diagrams for the GW (left), $T^{ph}$ 
(top-right) and $T^{pp}$ (bottom right) methods.}
\label{Diagrams1}
\end{figure}

\begin{table}[t!]
  \caption{\label{tab:1} Definitions of electronic two-particle tensors. The vertically
    grouped indices are combined into one (greek) super-index.}  \renewcommand{\arraystretch}{1.4}
  \begin{ruledtabular}
    \begin{tabular}{LLLL}
      \text{Quantity}& \text{2B and } GW & \multicolumn{1}{C}{T^{ph}} 
      & \multicolumn{1}{C}{T^{pp}}\\\hline
      \mcG^{ee}_{\begin{subarray}{c}13\\24\end{subarray}}&
      \mathcal{G}^{ee}_{1423}&
      \mathcal{G}^{ee}_{1432}&
      \mathcal{G}^{ee}_{1234}\\[1pt] 
      \mh^{e}_{\begin{subarray}{c}13\\24\end{subarray}}&
      h^{e}_{13}\delta_{42}-\delta_{13}h^{e}_{42}&
      h^{e}_{13}\delta_{42}-\delta_{13}h^{e}_{42}&
      h^{e}_{13}\delta_{24}+\delta_{13}h^{e}_{24}\\ 
      \mv_{\begin{subarray}{c}13\\24\end{subarray}} &
      v_{1432}&
      v_{1423}&
      v_{1243}\\[1pt]
      \mrho^{\lessgtr}_{\begin{subarray}{c}13\\24\end{subarray}}&
      \rho^{\lessgtr}_{13}\rho^{\gtrless}_{42}&
      \rho^{\lessgtr}_{13}\rho^{\gtrless}_{42}&
      \rho^{\lessgtr}_{13}\rho^{\lessgtr}_{24}\\
       \end{tabular}
    \end{ruledtabular}
\end{table}

In Fig.~\ref{Diagrams1}~(top) we show the diagrammatic representation of the 
two-particle GF $\cG^{ee}_{imjn}$ in 
the 2B$^{d}$ and 2B$^{x}$ approximation. They are obtained one from 
another by interchanging the external outgoing vertices $j$ and $n$. 
Alternatively, we can obtain one from another by exchanging the internal outoing (or incoming)  
vertices of the interaction line. The sum  
2B$^{d}+$2B$^{x}$  is usually named the second-Born (2B) approximation.
The two-particle GF in the $GW$, $T^{ph}$ and $T^{pp}$ 
approximation is illustrated in Fig.~\ref{Diagrams1}~(bottom). In 
Ref.~\cite{pavlyukh_photoinduced_2021} we proved that if one defines 
the matrices in the two-electron space as shown in Table~\ref{tab:1} then 
$\mcG^{ee(c)}$ satisfies the EOM
\begin{subequations}
\begin{align}
i\frac{d}{dt}\mcG^{ee(c)}&=-\mPsi^{e(c)}
+\mh^{e(c)}_{\rm eff}
\mcG^{ee(c)}-\mcG^{ee(c)}\mh^{e(c)\dagger}_{\rm eff},
\\
\mPsi^{e(c)}&=\mrho^{>(c)}\mv^{(c)}\mrho^{<(c)}-
\mrho^{<(c)}\mv^{(c)}\mrho^{>(c)},
\\
\mh^{e(c)}_{\rm eff}&=\mh^{e(c)}-\mrho^{\Delta(c)}\mv^{\prime(c)},
\end{align}
\label{EOMGeec}%
\end{subequations}
where the matrices $\mv^{(c)}$ and $\mv^{\prime(c)}$ are constructed from the four-index 
tensors 
\begin{subequations}
    \begin{align}
    v^{(c)}_{ijmn}&=a_{c}v_{ijmn}-b_{c}v_{ijnm},
    \\
    v^{\prime(c)}_{ijmn}&=a'_{c}v_{ijmn}-b'_{c}v_{ijnm}.
\end{align}    
\end{subequations}
\begin{table}[t!]
  \caption{\label{tab:2} Classes and parameters for all methods}  \renewcommand{\arraystretch}{1.4}
  \begin{ruledtabular}
    \begin{tabular}{LLLLLL}
      \text{Method}& {\rm class} & a_{c} & b_{c}  & a'_{c} & b'_{c} \\\hline
      {\rm 2B}^{d} & {\rm 2B} &  1  &  0  &  0  &  0 \\
      {\rm 2B}^{x} & {\rm 2B} &  0  &  1  &  0  &  0 \\ 
      GW           & GW &        1  &  0  &  1  &  0 \\
      GW^{x}       & GW &        0  &  1  &  1  &  0 \\
      XGW          & GW   &      0  &  1  &  0  &  1  \\
      GW+X         & GW   &      1  &  1  &  1  &  1  \\
      T^{ph}       & T^{ph} &    1  &  0  & -1  &  0 \\
      T^{phx}      & T^{ph} &    0  &  1  & -1  &  0 \\
      XT^{ph}      & T^{ph}   &  0  &  1  &  0  & -1  \\      
      T^{ph}+X     & T^{ph}   &  1  &  1  & -1  & -1  \\
      T^{pp}       & T^{pp} &    1  &  0  & -1  &  0 \\
      T^{ppx}      & T^{pp} &    0  &  1  & -1  &  0 \\
  \end{tabular}
    \end{ruledtabular}
\end{table}
In Table~\ref{tab:2} we report the values of 
$a_{c}$, $b_{c}$, $a'_{c}$, $b'_{c}$.
Notice that to the first order in $v$ the nonperturbative methods ($GW$, 
$T^{ph}$ and $T^{pp}$) reduce to 2B$^{d}$. Henceforth the matrices in 
the two electron space are constructed as illustrated in 
Table~\ref{tab:1} for all methods ``$c$'' belonging to the same 
``class'', see Table~\ref{tab:2}.

\begin{figure}[tbp]
\centering  \includegraphics[width=0.99\columnwidth]{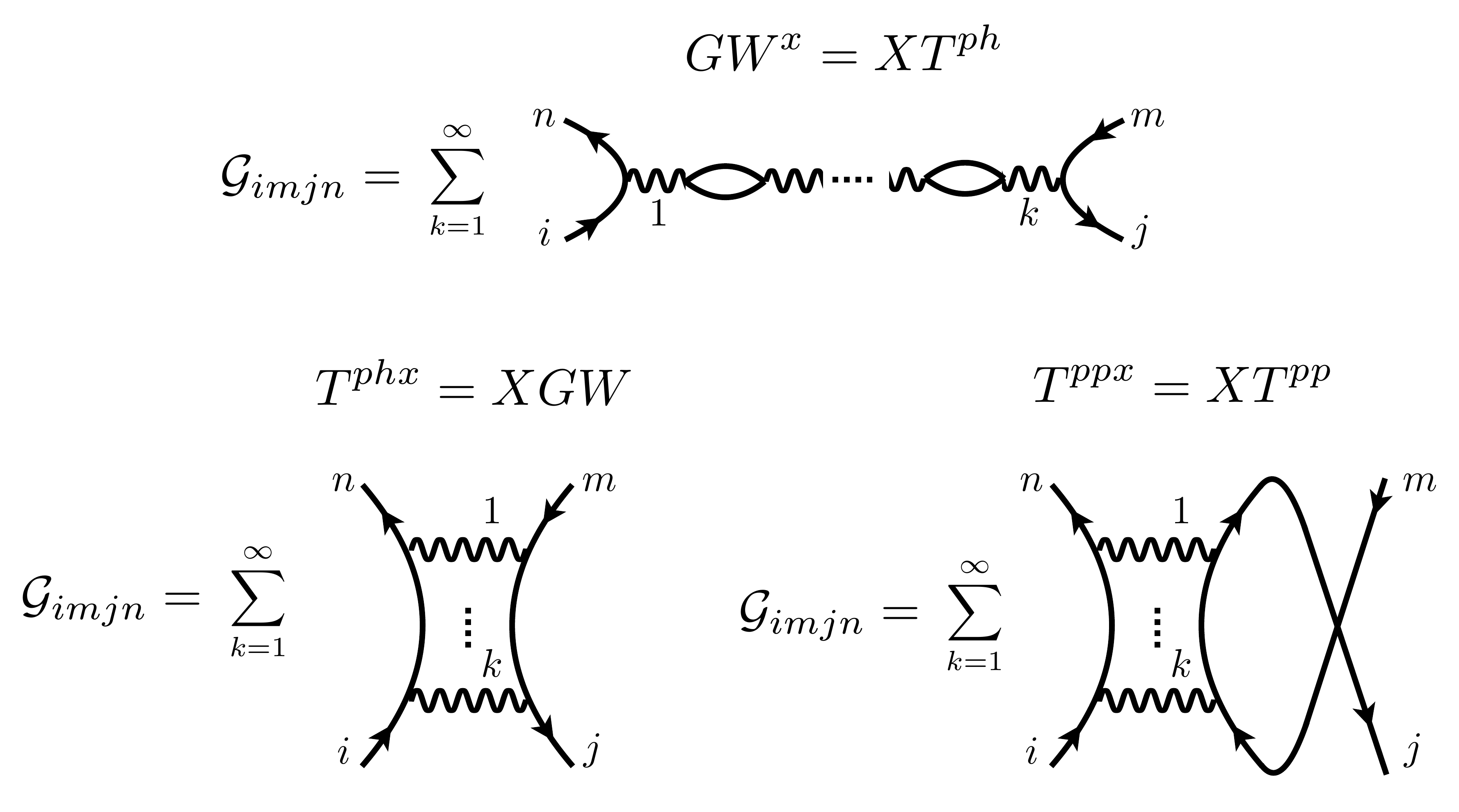}
\caption[]{Top: Diagrams for the $GW^{x}=XT^{ph}$ method. 
Bottom: Diagrams for the $T^{phx}=XGW$ (left) and $T^{ppx}$ 
(left)  methods.}
\label{Diagrams2}
\end{figure}

Exchange effects can be included in different ways. In analogy with 
the 2B method we could either exchange the outgoing vertices $j$ and $n$ 
or exchange the internal outgoing (or incoming) vertices of the 
interaction lines. 
Exchanging the incoming vertices $j$ and $n$ in $GW$, $T^{ph}$ and 
$T^{pp}$ leads to the  $GW^{x}$, $T^{phx}$ and $T^{ppx}$ 
approximations  illustrated in 
Fig.~\ref{Diagrams2}. Arranging the indices of the matrices according to the class 
these methods belong to ($GW^{x}$ like $GW$, $T^{phx}$ like $T^{ph}$ 
and $T^{ppx}$ like $T^{pp}$) we find again the EOM (\ref{EOMGeec}) 
with parameters given in Table~\ref{tab:2}. We observe that 
$h^{e(c)}_{\rm eff}$ is the same for the direct and exchange 
methods of the same class (same $a'_{c}$ and $b'_{c}$ parameters). 
This implies that if we are interested in treating correlations at 
the level of $2B=2B^{d}+2B^{x}$ or $GW+GW^{x}$ or $T^{ph}+T^{phx}$ or 
$T^{pp}+T^{ppx}$ we can sum the EOM for the direct and exchange 
methods, and propagate just one equation. The resulting EOM for the sum of the direct 
and exchange $\mcG^{ee(c)}$ is the same 
as the EOM of the only-direct or only-exchange method but $\mPsi^{e}$ is 
calculated with $a_{c}=b_{c}=1$.

Alternatively we can exchange the indices of the 
internal incoming (or outgoing) vertices of the interaction lines.
Graphically this exchange amounts to replace the 2B$^{d}$-like 
structures with the 2B$^{x}$ ones and {\em viceversa}.
If we apply this graphical rule to $GW$ we obtain the $XGW$ 
approximation which is identical to $T^{phx}$. Similarly, if  
we apply the  graphical rule to $T^{ph}$ we obtain the $XT^{ph}$ 
approximation which is identical to $GW^{x}$.
Arranging the indices like in $GW$ for $XGW$ and 
like in $T^{ph}$ for $XT^{ph}$ we find the EOM~(\ref{EOMGeec}) with 
parameters given in
Table~\ref{tab:2}.

The $T^{pp}$ diagrams behave differently. Under the exchange of the 
internal incoming (or outgoing) vertices of the interaction lines
a $T^{pp}$ diagram of order $n$ is mapped onto the same diagram if 
$n$ is even and onto the diagram of order $n$ of $T^{ppx}$ if $n$ is 
odd. Although this is a legitimate approximation it complicates the 
discussion on the double counting. We therefore do not address it 
further and write equivalently $T^{ppx}$ or $XT^{pp}$.

The inclusion of exchange effects like in $XGW$ and $XT^{ph}$ allows 
for constructing new approximations. If we  replace every 
interaction line $v_{ijmn}$ with the difference $(v_{ijmn}-v_{ijnm})$ 
then $GW\to GW+X$ and $T^{ph}\to T^{ph}+X$~\cite{pavlyukh_photoinduced_2021}.
Graphically this amounts to replace every 2B$^{d}$ structure
with the 2B$^{d}+$~2B$^{x}$ structure.
The $GW+X$ and $T^{ph}+X$ approximations solve the Bethe-Salpeter 
equation (BSE) with Hartree-Fock kernel in the two inequivalent 
particle-hole channels. The standard BSE used to calculate absorption 
spectra  corresponds to 
the $GW+X$ method~\cite{pavarini_linear_2016}. The EOM for these approximations are again given 
by Eq.~(\ref{EOMGeec}) with parameters given in
Table~\ref{tab:2}.

\section{How to combine different methods without double counting}
\label{appmerging}

We have seen in the previous Section that the index order of the 
matrices in Eq.~(\ref{EOMGee}) is common to all methods belonging to 
the same ``class'' (2B, 
$GW$, $T^{ph}$ or $T^{pp}$)~\cite{pavlyukh_photoinduced_2021}, 
and for $c$ in a given class 
the matrix elements of $\mv$ (appearing in $\mPsi^{e}$ and $\mh^{e}_{\rm 
eff}$) are calculated from the Coulomb tensor 
$v^{(c)}_{ijmn}=a_{c}v_{ijmn}-b_{c}v_{ijnm}$ (for $\mPsi^{e}$)
and $v^{\prime(c)}_{ijmn}=a'_{c}v_{ijmn}-b'_{c}v_{ijnm}$ (for $\mh^{e}_{\rm 
eff}$). The integers 
$a_{c},b_{c}$ and $a'_{c},b'_{c}$ take values between $-1$ and $1$,
see again Table~\ref{tab:2}. 

The most convenient way to avoid double countings is to treat 
the four integers  $n_{GW+X}$, $n_{T^{ph}+X}$, $n_{T^{pp}}$  and 
$n_{XT^{pp}}$ as independent and with values either 0 or 1. All other 
integers $n_{c}$ can then be chosen taking into account whether the 
method ``$c$'' is  already included. For instance if  $n_{GW+X}=1$ 
then $n_{GW}=-1,0$ whereas if $n_{GW+X}=0$ then  $n_{GW}=0,1$. 
We then have the following 
possibilities
\begin{align}
n_{GW}&=-n_{GW+X},1-n_{GW+X},
\nonumber \\
n_{XGW}&=-n_{GW+X},1-n_{GW+X},
\nonumber \\
n_{T^{ph}}&=-n_{T^{ph}+X},1-n_{T^{ph}+X},
\nonumber \\
n_{XT^{ph}}&= -n_{T^{ph}+X},1-n_{T^{ph}+X}.
\nonumber
\end{align}
The possible values of $n_{{\rm 2B}^{d}}$ can instead be $-N_{d},1-N_{d}$ where $N_{d}$ is the number 
of times that the second-order direct term is included:
$N_{d}=n_{GW}+n_{GW+X}+n_{T^{ph}}+n_{T^{ph}+X}+n_{T^{pp}}$.
Similarly $n_{{\rm 2B}^{x}}=-N_{x},1-N_{x}$ where $N_{x}$ is the number 
of times that the second-order exchange term is included:
$N_{x}=n_{XGW}+n_{GW+X}+n_{XT^{ph}}+n_{T^{ph}+X}+n_{XT^{pp}}$.

\section{Numerical details}
\label{appnum}

To isolate the correlation-induced charge migration of the Gly I 
conformer resulting from 
the removal of an electron from the $12a'$ MO it is sufficient to 
consider the four MOs $12a'$ (HOMO-8), $4a''$ (HOMO-2), $16a'$ (HOMO) 
and $5a''$ 
(LUMO)~\cite{kuleff_multielectron_2005,kuleff_charge_2007,cooper_single-photon_2013,perfetto_first-principles_2019,pavlyukh_photoinduced_2021}. 
Freezing all other electrons and working in the Hartree-Fock (HF) MO basis 
the electronic Hamiltonian in second quantization reads
\begin{align}
\hat{H}_{\rm el}&=\sum_{ij\s}\big(\delta_{ij}\epsilon^{\rm 
HF}_{i}-V^{\rm HF}_{ij}\big)\hat{d}^{\dag}_{i\s}\hat{d}_{j\s}
\nonumber \\
&+\frac{1}{2}\sum_{ijmn}\sum_{\s\s'}v_{ijmn}\hat{d}^{\dag}_{i\s}\hat{d}^{\dag}_{j\s'}
\hat{d}_{m\s'}\hat{d}_{n\s},
\label{hamel}
\end{align}
where $\epsilon^{\rm HF}_{i}=(-0.704, -0.475, -0.400, 0.176)$~a.u. are the HF single-particle energies of 
the neutral molecule and $V^{\rm HF}$ is the HF potential generated by 
the active electrons; the sums run over spin and the four MOs. 
The shake-up process is activated by the Coulomb integral 
$v_{16a'4a''5a''12a'}=0.017$~a.u. and other integrals connected to it by the 
symmetry relations (for real MOs)
\begin{align}
v_{ijmn}=v_{imjn}=v_{njmi}=v_{jinm}.
\label{symmrel}
\end{align}
The renormalization of the energy of the shake-up state is instead mainly due to 
the direct integral 
$v^{d}_{4a''5a''}\equiv v_{4a''5a''5a''4a''}=0.39$~a.u., 
exchange integral $v^{x}_{4a''5a''}\equiv v_{4a''5a''4a''5a''}=0.08$~a.u. and all 
other integrals connected to these two through the symmetry relations 
in Eq.~(\ref{symmrel}). The renormalization due to $v^{d}_{4a''5a''}$ 
is captured by the $XGW$ 
approximation whereas the 
renormalization due $v^{x}_{4a''5a''}$ is captured by the $GW$ 
approximation~\cite{pavlyukh_photoinduced_2021}. To simplify the 
discussion we have discarded $v^{d}_{4a''5a''}$; no complication 
arises in adding exchange to the $G\tilde{W}$ method. 

To describe the molecule in a cavity we add to the reduced 
electronic Hamiltonian in Eq.~(\ref{hamel}) the free-photon 
Hamiltonian and the electron-photon interaction
\begin{align}
\hat{H}_{\rm 
cavity}=\omega_{0}(\hat{a}^{\dag}\hat{a}+\frac{1}{2})+\sum_{ij\s}\frac{g_{ij}}{\sqrt{2}}\hat{d}^{\dag}_{i\s}\hat{d}_{j\s}
(\hat{a}^{\dag}+\hat{a}).
\end{align}
We study the case of a cavity-photon coupled to the transition 
$\Psi_{\rm shake-up}\to\Psi_{\rm cation}$ and therefore choose 
$g_{ij}=g_{ji}=g\neq 0$ only for the pair $4a''$ and $5a''$ of  MOs.   
As detailed in the main text
$g=\lambda d_{4a'',5a''}\sqrt{\omega_{0}}$, where 
$d_{4a'',5a''}=0.125$~a.u. is the dipole moment (averaged over three orthogonal 
directions) and  $\lambda=0.212$~a.u.. is the
electron-photon coupling strength~\cite{yang_quantum_2021}.

In CI we first calculate the ground state $\Psi_{g}$ of 
the molecule in the cavity. At convergence the number of photons 
$n_{\rm ph}=\langle\Psi_{g}|\hat{a}^{\dag}\hat{a}|\Psi_{g}\rangle$ is of 
the order of $10^{-4}$, consistent with the fact that cavity-photons are 
emitted only in the transition between {\em cationic states}. To ionize 
the molecule from the $12a'$ MO we couple this state to a fictitious 
vacuum state 
\begin{align}
    \hat{H}_{\rm laser}(t)=\sum_{\s}R(t)\,(\hat{d}^{\dag}_{12a'\s}\hat{d}_{\rm vacuum\,\s}+{\rm h.c.})
\end{align}
where the Rabi coupling 
\begin{align}
 R(t)=R_{0}\theta(t)\theta(\tau-t)\sin^{2}(\frac{\pi 
t}{\tau})\sin(\omega_{\rm laser}t)
\end{align}
describes a laser pulse of duration $\tau$ centered at frequency 
$\omega_{\rm laser}$. The intensity $R_{0}$ is chosen small enough to work in the 
linear response regime, hence we check that the population of the $12a'$ MO just 
after the pulse satisfies $\delta n_{12a'}\equiv n_{12a'}(\tau)-n_{12a'}(0)={\mathcal O}(R_{0}^{2})$. We 
solve the time-dependent Schr\"odinger equation
\begin{align}
i\frac{d}{dt}|\Psi(t)\rangle=\big(\hat{H}_{\rm el}+\hat{H}_{\rm cavity}+
 \hat{H}_{\rm laser}(t)\big)|\Psi(t)\rangle
 \end{align}
with initial condition $|\Psi(t)\rangle=|\Psi_{g}\rangle$ for 
different photon frequencies $\omega_{0}$. In Fig.~\ref{gedanken}(d) we 
show the Fourier transform of $n_{12a'}(t)=\sum_{\s}
\langle\Psi(t)|\hat{d}^{\dag}_{12a'\s}\hat{d}_{12a'\s}|\Psi(t)\rangle$.

In the GKBA+ODE we use the fact that the ground state $\Psi_{g}$ is 
weakly correlated and we approximate it with the HF ground-state with 
no photons. How to discard initial correlations in GKBA+ODE has 
already been discussed in Ref.~\cite{pavlyukh_photoinduced_2021}. In 
short this is done by calculating the electronic driving 
$\mPsi^{e}(t)$ defined in Eq.~(\ref{psie}) using only the shake-up 
Coulomb integrals and by setting to zero the bosonic driving 
$\mPsi^{b}(t)$ defined in Eq.~(\ref{Psi:b:def}). The initial 
conditions for the bosonic displacements and density matrix 
describing an initial state with no photons are
\begin{align}
\phi_{\xi}(0)=0,\quad\quad
\gamma_{\xi\xi'}(0)=\frac{1}{2}\left(\begin{array}{cc}
1 & -i \\ i & 1 \end{array}\right)_{\xi\xi'}.
\end{align}
The initial condition for the electronic density matrix describing 
the photoionized molecule from the $12a'$ MO is taken as 
\begin{align}
\rho(0)={\rm diag}(1-\frac{\delta n_{12a'}}{2},2,2,0),
\end{align}
where $\delta n_{12a'}$ is the depopulation obtained from the CI 
calculation. The initial condition for the high order GFs is simply $\mcG^{ee}=\mcG^{eb}=\mcG^{b}=\mcA=0$. It is 
straightforward to verify that for $\delta n_{12a'}=0$ this set of 
initial conditions are a stationary solution of the GKBA+ODE 
equations for all methods.
In Fig.~\ref{gedanken}(d) we 
show the Fourier transform of $n_{12a'}(t)=\rho_{12a'12a'}(t)$ in 
three different diagrammatic approximations.

%\bibliography{MyLibrary,Amendments}

\begin{thebibliography}{59}%
\makeatletter
\providecommand \@ifxundefined [1]{%
 \@ifx{#1\undefined}
}%
\providecommand \@ifnum [1]{%
 \ifnum #1\expandafter \@firstoftwo
 \else \expandafter \@secondoftwo
 \fi
}%
\providecommand \@ifx [1]{%
 \ifx #1\expandafter \@firstoftwo
 \else \expandafter \@secondoftwo
 \fi
}%
\providecommand \natexlab [1]{#1}%
\providecommand \enquote  [1]{``#1''}%
\providecommand \bibnamefont  [1]{#1}%
\providecommand \bibfnamefont [1]{#1}%
\providecommand \citenamefont [1]{#1}%
\providecommand \href@noop [0]{\@secondoftwo}%
\providecommand \href [0]{\begingroup \@sanitize@url \@href}%
\providecommand \@href[1]{\@@startlink{#1}\@@href}%
\providecommand \@@href[1]{\endgroup#1\@@endlink}%
\providecommand \@sanitize@url [0]{\catcode `\\12\catcode `\$12\catcode
  `\&12\catcode `\#12\catcode `\^12\catcode `\_12\catcode `\%12\relax}%
\providecommand \@@startlink[1]{}%
\providecommand \@@endlink[0]{}%
\providecommand \url  [0]{\begingroup\@sanitize@url \@url }%
\providecommand \@url [1]{\endgroup\@href {#1}{\urlprefix }}%
\providecommand \urlprefix  [0]{URL }%
\providecommand \Eprint [0]{\href }%
\providecommand \doibase [0]{http://dx.doi.org/}%
\providecommand \selectlanguage [0]{\@gobble}%
\providecommand \bibinfo  [0]{\@secondoftwo}%
\providecommand \bibfield  [0]{\@secondoftwo}%
\providecommand \translation [1]{[#1]}%
\providecommand \BibitemOpen [0]{}%
\providecommand \bibitemStop [0]{}%
\providecommand \bibitemNoStop [0]{.\EOS\space}%
\providecommand \EOS [0]{\spacefactor3000\relax}%
\providecommand \BibitemShut  [1]{\csname bibitem#1\endcsname}%
\let\auto@bib@innerbib\@empty
%</preamble>
\bibitem [{\citenamefont {Feynman}(1949)}]{feynman_space-time_1949}%
  \BibitemOpen
  \bibfield  {author} {\bibinfo {author} {\bibfnamefont {R.~P.}\ \bibnamefont
  {Feynman}},\ }\href {\doibase 10.1103/PhysRev.76.769} {\bibfield  {journal}
  {\bibinfo  {journal} {Phys. Rev.}\ }\textbf {\bibinfo {volume} {76}},\
  \bibinfo {pages} {769} (\bibinfo {year} {1949})}\BibitemShut {NoStop}%
\bibitem [{\citenamefont {Abrikosov}\ \emph {et~al.}(1975)\citenamefont
  {Abrikosov}, \citenamefont {Gor'kov},\ and\ \citenamefont
  {Dzialoshinskii}}]{abrikosov_methods_1975}%
  \BibitemOpen
  \bibfield  {author} {\bibinfo {author} {\bibfnamefont {A.~A.}\ \bibnamefont
  {Abrikosov}}, \bibinfo {author} {\bibfnamefont {L.~P.}\ \bibnamefont
  {Gor'kov}}, \ and\ \bibinfo {author} {\bibfnamefont {I.~E.}\ \bibnamefont
  {Dzialoshinskii}},\ }\href@noop {} {\emph {\bibinfo {title} {Methods of
  quantum field theory in statistical physics}}}\ (\bibinfo  {publisher} {Dover
  Publications},\ \bibinfo {address} {New York},\ \bibinfo {year}
  {1975})\BibitemShut {NoStop}%
\bibitem [{\citenamefont {Mattuck}(1992)}]{mattuck_guide_1992}%
  \BibitemOpen
  \bibfield  {author} {\bibinfo {author} {\bibfnamefont {R.~D.}\ \bibnamefont
  {Mattuck}},\ }\href@noop {} {\emph {\bibinfo {title} {A guide to {Feynman}
  diagrams in the many-body problem}}},\ \bibinfo {edition} {2nd}\ ed.\
  (\bibinfo  {publisher} {Dover Publications},\ \bibinfo {address} {New York},\
  \bibinfo {year} {1992})\BibitemShut {NoStop}%
\bibitem [{\citenamefont {Fetter}\ and\ \citenamefont
  {Walecka}(2003)}]{fetter_quantum_2003}%
  \BibitemOpen
  \bibfield  {author} {\bibinfo {author} {\bibfnamefont {A.}~\bibnamefont
  {Fetter}}\ and\ \bibinfo {author} {\bibfnamefont {J.}~\bibnamefont
  {Walecka}},\ }\href {http://books.google.com/books?id=0wekf1s83b0C} {\emph
  {\bibinfo {title} {Quantum theory of many-particle systems}}},\ Dover {Books}
  on {Physics}\ (\bibinfo  {publisher} {Dover Publications},\ \bibinfo {year}
  {2003})\BibitemShut {NoStop}%
\bibitem [{\citenamefont {Gross}\ \emph {et~al.}(1991)\citenamefont {Gross},
  \citenamefont {Runge},\ and\ \citenamefont
  {Heinonen}}]{gross_many-particle_1991}%
  \BibitemOpen
  \bibfield  {author} {\bibinfo {author} {\bibfnamefont {E.~K.~U.}\
  \bibnamefont {Gross}}, \bibinfo {author} {\bibfnamefont {E.}~\bibnamefont
  {Runge}}, \ and\ \bibinfo {author} {\bibfnamefont {O.}~\bibnamefont
  {Heinonen}},\ }\href@noop {} {\emph {\bibinfo {title} {Many-particle
  theory}}}\ (\bibinfo  {publisher} {A. Hilger},\ \bibinfo {year}
  {1991})\BibitemShut {NoStop}%
\bibitem [{\citenamefont {Konstantinov}\ and\ \citenamefont
  {Perel}(1961)}]{konstantinov_diagram_1961}%
  \BibitemOpen
  \bibfield  {author} {\bibinfo {author} {\bibfnamefont {O.~V.}\ \bibnamefont
  {Konstantinov}}\ and\ \bibinfo {author} {\bibfnamefont {V.~I.}\ \bibnamefont
  {Perel}},\ }\href@noop {} {\bibfield  {journal} {\bibinfo  {journal} {Sov.
  Phys. JETP}\ }\textbf {\bibinfo {volume} {12}},\ \bibinfo {pages} {142}
  (\bibinfo {year} {1961})}\BibitemShut {NoStop}%
\bibitem [{\citenamefont {Keldysh}(1965)}]{keldysh_diagram_1965}%
  \BibitemOpen
  \bibfield  {author} {\bibinfo {author} {\bibfnamefont {L.~V.}\ \bibnamefont
  {Keldysh}},\ }\href@noop {} {\bibfield  {journal} {\bibinfo  {journal} {Sov.
  Phys. JETP}\ }\textbf {\bibinfo {volume} {20}},\ \bibinfo {pages} {1018}
  (\bibinfo {year} {1965})}\BibitemShut {NoStop}%
\bibitem [{\citenamefont {Kadanoff}\ and\ \citenamefont
  {Baym}(1962)}]{kadanoff_quantum_1962}%
  \BibitemOpen
  \bibfield  {author} {\bibinfo {author} {\bibfnamefont {L.}~\bibnamefont
  {Kadanoff}}\ and\ \bibinfo {author} {\bibfnamefont {G.}~\bibnamefont
  {Baym}},\ }\href@noop {} {\emph {\bibinfo {title} {Quantum statistical
  mechanics {Green}'s function methods in equilibrium and nonequilibrium
  problems}}}\ (\bibinfo  {publisher} {W.A. Benjamin},\ \bibinfo {address} {New
  York},\ \bibinfo {year} {1962})\BibitemShut {NoStop}%
\bibitem [{\citenamefont {Stefanucci}\ and\ \citenamefont {van
  Leeuwen}(2013)}]{stefanucci_nonequilibrium_2013}%
  \BibitemOpen
  \bibfield  {author} {\bibinfo {author} {\bibfnamefont {G.}~\bibnamefont
  {Stefanucci}}\ and\ \bibinfo {author} {\bibfnamefont {R.}~\bibnamefont {van
  Leeuwen}},\ }\href {http://dx.doi.org/10.1017/CBO9781139023979} {\emph
  {\bibinfo {title} {Nonequilibrium {Many}-{Body} {Theory} of {Quantum}
  {Systems}: {A} {Modern} {Introduction}}}}\ (\bibinfo  {publisher} {Cambridge
  University Press},\ \bibinfo {address} {Cambridge},\ \bibinfo {year}
  {2013})\BibitemShut {NoStop}%
\bibitem [{\citenamefont {Stan}\ \emph {et~al.}(2009)\citenamefont {Stan},
  \citenamefont {Dahlen},\ and\ \citenamefont {van Leeuwen}}]{stan_time_2009}%
  \BibitemOpen
  \bibfield  {author} {\bibinfo {author} {\bibfnamefont {A.}~\bibnamefont
  {Stan}}, \bibinfo {author} {\bibfnamefont {N.~E.}\ \bibnamefont {Dahlen}}, \
  and\ \bibinfo {author} {\bibfnamefont {R.}~\bibnamefont {van Leeuwen}},\
  }\href {\doibase 10.1063/1.3127247} {\bibfield  {journal} {\bibinfo
  {journal} {J. Chem. Phys.}\ }\textbf {\bibinfo {volume} {130}},\ \bibinfo
  {pages} {224101} (\bibinfo {year} {2009})}\BibitemShut {NoStop}%
\bibitem [{\citenamefont {Balzer}\ and\ \citenamefont
  {Bonitz}(2013)}]{balzer_nonequilibrium_2013-1}%
  \BibitemOpen
  \bibfield  {author} {\bibinfo {author} {\bibfnamefont {K.}~\bibnamefont
  {Balzer}}\ and\ \bibinfo {author} {\bibfnamefont {M.}~\bibnamefont
  {Bonitz}},\ }\href@noop {} {\emph {\bibinfo {title} {Nonequilibrium {Green}'s
  function approach to inhomogeneous systems}}},\ \bibinfo {series} {Lecture
  notes in physics}\ No.\ \bibinfo {number} {867}\ (\bibinfo  {publisher}
  {Springer},\ \bibinfo {address} {Heidelberg},\ \bibinfo {year}
  {2013})\BibitemShut {NoStop}%
\bibitem [{\citenamefont {Sch\"{u}ler}\ \emph {et~al.}(2016)\citenamefont
  {Sch\"{u}ler}, \citenamefont {Berakdar},\ and\ \citenamefont
  {Pavlyukh}}]{schuler_time-dependent_2016}%
  \BibitemOpen
  \bibfield  {author} {\bibinfo {author} {\bibfnamefont {M.}~\bibnamefont
  {Sch\"{u}ler}}, \bibinfo {author} {\bibfnamefont {J.}~\bibnamefont
  {Berakdar}}, \ and\ \bibinfo {author} {\bibfnamefont {Y.}~\bibnamefont
  {Pavlyukh}},\ }\href {\doibase 10.1103/PhysRevB.93.054303} {\bibfield
  {journal} {\bibinfo  {journal} {Phys. Rev. B}\ }\textbf {\bibinfo {volume}
  {93}},\ \bibinfo {pages} {054303} (\bibinfo {year} {2016})}\BibitemShut
  {NoStop}%
\bibitem [{\citenamefont {Sch\"{u}ler}\ \emph {et~al.}(2020)\citenamefont
  {Sch\"{u}ler}, \citenamefont {Gole\v{z}}, \citenamefont {Murakami},
  \citenamefont {Bittner}, \citenamefont {Herrmann}, \citenamefont {Strand},
  \citenamefont {Werner},\ and\ \citenamefont {Eckstein}}]{schuler_nessi_2020}%
  \BibitemOpen
  \bibfield  {author} {\bibinfo {author} {\bibfnamefont {M.}~\bibnamefont
  {Sch\"{u}ler}}, \bibinfo {author} {\bibfnamefont {D.}~\bibnamefont
  {Gole\v{z}}}, \bibinfo {author} {\bibfnamefont {Y.}~\bibnamefont {Murakami}},
  \bibinfo {author} {\bibfnamefont {N.}~\bibnamefont {Bittner}}, \bibinfo
  {author} {\bibfnamefont {A.}~\bibnamefont {Herrmann}}, \bibinfo {author}
  {\bibfnamefont {H.~U.}\ \bibnamefont {Strand}}, \bibinfo {author}
  {\bibfnamefont {P.}~\bibnamefont {Werner}}, \ and\ \bibinfo {author}
  {\bibfnamefont {M.}~\bibnamefont {Eckstein}},\ }\href {\doibase
  10.1016/j.cpc.2020.107484} {\bibfield  {journal} {\bibinfo  {journal} {Comp.
  Phys. Commun.}\ }\textbf {\bibinfo {volume} {257}},\ \bibinfo {pages}
  {107484} (\bibinfo {year} {2020})}\BibitemShut {NoStop}%
\bibitem [{\citenamefont {Pavlyukh}\ \emph
  {et~al.}(2022{\natexlab{a}})\citenamefont {Pavlyukh}, \citenamefont
  {Perfetto}, \citenamefont {Karlsson}, \citenamefont {van Leeuwen},\ and\
  \citenamefont {Stefanucci}}]{pavlyukh_time-linear_2021}%
  \BibitemOpen
  \bibfield  {author} {\bibinfo {author} {\bibfnamefont {Y.}~\bibnamefont
  {Pavlyukh}}, \bibinfo {author} {\bibfnamefont {E.}~\bibnamefont {Perfetto}},
  \bibinfo {author} {\bibfnamefont {D.}~\bibnamefont {Karlsson}}, \bibinfo
  {author} {\bibfnamefont {R.}~\bibnamefont {van Leeuwen}}, \ and\ \bibinfo
  {author} {\bibfnamefont {G.}~\bibnamefont {Stefanucci}},\ }\href@noop {} {}
  (\bibinfo {year} {2022}{\natexlab{a}}),\ \bibinfo {note} {{Phys. Rev. B}, in
  press [arXiv:2111.06698]}\BibitemShut {NoStop}%
\bibitem [{\citenamefont {Lipavský}\ \emph {et~al.}(1986)\citenamefont
  {Lipavský}, \citenamefont {\v{S}pi\v{c}ka},\ and\ \citenamefont
  {Velický}}]{lipavsky_generalized_1986}%
  \BibitemOpen
  \bibfield  {author} {\bibinfo {author} {\bibfnamefont {P.}~\bibnamefont
  {Lipavský}}, \bibinfo {author} {\bibfnamefont {V.}~\bibnamefont
  {\v{S}pi\v{c}ka}}, \ and\ \bibinfo {author} {\bibfnamefont {B.}~\bibnamefont
  {Velický}},\ }\href {\doibase 10.1103/PhysRevB.34.6933} {\bibfield
  {journal} {\bibinfo  {journal} {Phys. Rev. B}\ }\textbf {\bibinfo {volume}
  {34}},\ \bibinfo {pages} {6933} (\bibinfo {year} {1986})}\BibitemShut
  {NoStop}%
\bibitem [{\citenamefont {Karlsson}\ \emph {et~al.}(2021)\citenamefont
  {Karlsson}, \citenamefont {van Leeuwen}, \citenamefont {Pavlyukh},
  \citenamefont {Perfetto},\ and\ \citenamefont
  {Stefanucci}}]{karlsson_fast_2021}%
  \BibitemOpen
  \bibfield  {author} {\bibinfo {author} {\bibfnamefont {D.}~\bibnamefont
  {Karlsson}}, \bibinfo {author} {\bibfnamefont {R.}~\bibnamefont {van
  Leeuwen}}, \bibinfo {author} {\bibfnamefont {Y.}~\bibnamefont {Pavlyukh}},
  \bibinfo {author} {\bibfnamefont {E.}~\bibnamefont {Perfetto}}, \ and\
  \bibinfo {author} {\bibfnamefont {G.}~\bibnamefont {Stefanucci}},\ }\href
  {\doibase 10.1103/PhysRevLett.127.036402} {\bibfield  {journal} {\bibinfo
  {journal} {Phys. Rev. Lett.}\ }\textbf {\bibinfo {volume} {127}},\ \bibinfo
  {pages} {036402} (\bibinfo {year} {2021})}\BibitemShut {NoStop}%
\bibitem [{\citenamefont {Schl\"{u}nzen}\ \emph {et~al.}(2020)\citenamefont
  {Schl\"{u}nzen}, \citenamefont {Joost},\ and\ \citenamefont
  {Bonitz}}]{schlunzen_achieving_2020}%
  \BibitemOpen
  \bibfield  {author} {\bibinfo {author} {\bibfnamefont {N.}~\bibnamefont
  {Schl\"{u}nzen}}, \bibinfo {author} {\bibfnamefont {J.-P.}\ \bibnamefont
  {Joost}}, \ and\ \bibinfo {author} {\bibfnamefont {M.}~\bibnamefont
  {Bonitz}},\ }\href {\doibase 10.1103/PhysRevLett.124.076601} {\bibfield
  {journal} {\bibinfo  {journal} {Phys. Rev. Lett.}\ }\textbf {\bibinfo
  {volume} {124}},\ \bibinfo {pages} {076601} (\bibinfo {year}
  {2020})}\BibitemShut {NoStop}%
\bibitem [{\citenamefont {Joost}\ \emph {et~al.}(2020)\citenamefont {Joost},
  \citenamefont {Schl\"{u}nzen},\ and\ \citenamefont
  {Bonitz}}]{joost_g1-g2_2020}%
  \BibitemOpen
  \bibfield  {author} {\bibinfo {author} {\bibfnamefont {J.-P.}\ \bibnamefont
  {Joost}}, \bibinfo {author} {\bibfnamefont {N.}~\bibnamefont
  {Schl\"{u}nzen}}, \ and\ \bibinfo {author} {\bibfnamefont {M.}~\bibnamefont
  {Bonitz}},\ }\href {\doibase 10.1103/PhysRevB.101.245101} {\bibfield
  {journal} {\bibinfo  {journal} {Phys. Rev. B}\ }\textbf {\bibinfo {volume}
  {101}},\ \bibinfo {pages} {245101} (\bibinfo {year} {2020})}\BibitemShut
  {NoStop}%
\bibitem [{\citenamefont {Perfetto}\ \emph {et~al.}(2022)\citenamefont
  {Perfetto}, \citenamefont {Pavlyukh},\ and\ \citenamefont
  {Stefanucci}}]{perfetto_real-time_2022}%
  \BibitemOpen
  \bibfield  {author} {\bibinfo {author} {\bibfnamefont {E.}~\bibnamefont
  {Perfetto}}, \bibinfo {author} {\bibfnamefont {Y.}~\bibnamefont {Pavlyukh}},
  \ and\ \bibinfo {author} {\bibfnamefont {G.}~\bibnamefont {Stefanucci}},\
  }\href {\doibase 10.1103/PhysRevLett.128.016801} {\bibfield  {journal}
  {\bibinfo  {journal} {Phys. Rev. Lett.}\ }\textbf {\bibinfo {volume} {128}},\
  \bibinfo {pages} {016801} (\bibinfo {year} {2022})}\BibitemShut {NoStop}%
\bibitem [{\citenamefont {Pavlyukh}\ \emph {et~al.}(2021)\citenamefont
  {Pavlyukh}, \citenamefont {Perfetto},\ and\ \citenamefont
  {Stefanucci}}]{pavlyukh_photoinduced_2021}%
  \BibitemOpen
  \bibfield  {author} {\bibinfo {author} {\bibfnamefont {Y.}~\bibnamefont
  {Pavlyukh}}, \bibinfo {author} {\bibfnamefont {E.}~\bibnamefont {Perfetto}},
  \ and\ \bibinfo {author} {\bibfnamefont {G.}~\bibnamefont {Stefanucci}},\
  }\href {\doibase 10.1103/PhysRevB.104.035124} {\bibfield  {journal} {\bibinfo
   {journal} {Phys. Rev. B}\ }\textbf {\bibinfo {volume} {104}},\ \bibinfo
  {pages} {035124} (\bibinfo {year} {2021})}\BibitemShut {NoStop}%
\bibitem [{\citenamefont {Frederiksen}\ \emph {et~al.}(2007)\citenamefont
  {Frederiksen}, \citenamefont {Paulsson}, \citenamefont {Brandbyge},\ and\
  \citenamefont {Jauho}}]{frederiksen_inelastic_2007}%
  \BibitemOpen
  \bibfield  {author} {\bibinfo {author} {\bibfnamefont {T.}~\bibnamefont
  {Frederiksen}}, \bibinfo {author} {\bibfnamefont {M.}~\bibnamefont
  {Paulsson}}, \bibinfo {author} {\bibfnamefont {M.}~\bibnamefont {Brandbyge}},
  \ and\ \bibinfo {author} {\bibfnamefont {A.-P.}\ \bibnamefont {Jauho}},\
  }\href {\doibase 10.1103/PhysRevB.75.205413} {\bibfield  {journal} {\bibinfo
  {journal} {Phys. Rev. B}\ }\textbf {\bibinfo {volume} {75}},\ \bibinfo
  {pages} {205413} (\bibinfo {year} {2007})}\BibitemShut {NoStop}%
\bibitem [{\citenamefont {Cannuccia}\ and\ \citenamefont
  {Marini}(2011)}]{cannuccia_effect_2011}%
  \BibitemOpen
  \bibfield  {author} {\bibinfo {author} {\bibfnamefont {E.}~\bibnamefont
  {Cannuccia}}\ and\ \bibinfo {author} {\bibfnamefont {A.}~\bibnamefont
  {Marini}},\ }\href {\doibase 10.1103/PhysRevLett.107.255501} {\bibfield
  {journal} {\bibinfo  {journal} {Phys. Rev. Lett.}\ }\textbf {\bibinfo
  {volume} {107}},\ \bibinfo {pages} {255501} (\bibinfo {year}
  {2011})}\BibitemShut {NoStop}%
\bibitem [{\citenamefont {Pavlyukh}\ \emph
  {et~al.}(2022{\natexlab{b}})\citenamefont {Pavlyukh}, \citenamefont
  {Perfetto}, \citenamefont {Karlsson}, \citenamefont {van Leeuwen},\ and\
  \citenamefont {Stefanucci}}]{pavlyukh_time-linearII_2021}%
  \BibitemOpen
  \bibfield  {author} {\bibinfo {author} {\bibfnamefont {Y.}~\bibnamefont
  {Pavlyukh}}, \bibinfo {author} {\bibfnamefont {E.}~\bibnamefont {Perfetto}},
  \bibinfo {author} {\bibfnamefont {D.}~\bibnamefont {Karlsson}}, \bibinfo
  {author} {\bibfnamefont {R.}~\bibnamefont {van Leeuwen}}, \ and\ \bibinfo
  {author} {\bibfnamefont {G.}~\bibnamefont {Stefanucci}},\ }\href@noop {} {}
  (\bibinfo {year} {2022}{\natexlab{b}}),\ \bibinfo {note} {{Phys. Rev. B}, in
  press [arXiv:2111.06699]}\BibitemShut {NoStop}%
\bibitem [{\citenamefont {Rizzi}\ \emph {et~al.}(2016)\citenamefont {Rizzi},
  \citenamefont {Todorov}, \citenamefont {Kohanoff},\ and\ \citenamefont
  {Correa}}]{rizzi_electron-phonon_2016}%
  \BibitemOpen
  \bibfield  {author} {\bibinfo {author} {\bibfnamefont {V.}~\bibnamefont
  {Rizzi}}, \bibinfo {author} {\bibfnamefont {T.~N.}\ \bibnamefont {Todorov}},
  \bibinfo {author} {\bibfnamefont {J.~J.}\ \bibnamefont {Kohanoff}}, \ and\
  \bibinfo {author} {\bibfnamefont {A.~A.}\ \bibnamefont {Correa}},\ }\href
  {\doibase 10.1103/PhysRevB.93.024306} {\bibfield  {journal} {\bibinfo
  {journal} {Phys. Rev. B}\ }\textbf {\bibinfo {volume} {93}},\ \bibinfo
  {pages} {024306} (\bibinfo {year} {2016})}\BibitemShut {NoStop}%
\bibitem [{\citenamefont {Baym}\ and\ \citenamefont
  {Kadanoff}(1961)}]{baym_conservation_1961}%
  \BibitemOpen
  \bibfield  {author} {\bibinfo {author} {\bibfnamefont {G.}~\bibnamefont
  {Baym}}\ and\ \bibinfo {author} {\bibfnamefont {L.~P.}\ \bibnamefont
  {Kadanoff}},\ }\href {\doibase 10.1103/PhysRev.124.287} {\bibfield  {journal}
  {\bibinfo  {journal} {Phys. Rev.}\ }\textbf {\bibinfo {volume} {124}},\
  \bibinfo {pages} {287} (\bibinfo {year} {1961})}\BibitemShut {NoStop}%
\bibitem [{\citenamefont {Baym}(1962)}]{baym_self-consistent_1962}%
  \BibitemOpen
  \bibfield  {author} {\bibinfo {author} {\bibfnamefont {G.}~\bibnamefont
  {Baym}},\ }\href {\doibase 10.1103/PhysRev.127.1391} {\bibfield  {journal}
  {\bibinfo  {journal} {Phys. Rev.}\ }\textbf {\bibinfo {volume} {127}},\
  \bibinfo {pages} {1391} (\bibinfo {year} {1962})}\BibitemShut {NoStop}%
\bibitem [{\citenamefont {van
  Leeuwen}(2004)}]{van_leeuwen_first-principles_2004}%
  \BibitemOpen
  \bibfield  {author} {\bibinfo {author} {\bibfnamefont {R.}~\bibnamefont {van
  Leeuwen}},\ }\href {\doibase 10.1103/PhysRevB.69.115110} {\bibfield
  {journal} {\bibinfo  {journal} {Phys. Rev. B}\ }\textbf {\bibinfo {volume}
  {69}},\ \bibinfo {pages} {115110} (\bibinfo {year} {2004})}\BibitemShut
  {NoStop}%
\bibitem [{\citenamefont {Karlsson}\ and\ \citenamefont {van
  Leeuwen}(2020)}]{andreoni_non-equilibrium_2020}%
  \BibitemOpen
  \bibfield  {author} {\bibinfo {author} {\bibfnamefont {D.}~\bibnamefont
  {Karlsson}}\ and\ \bibinfo {author} {\bibfnamefont {R.}~\bibnamefont {van
  Leeuwen}},\ }in\ \href {\doibase 10.1007/978-3-319-44677-6_8} {\emph
  {\bibinfo {booktitle} {Handbook of {Materials} {Modeling}}}},\ \bibinfo
  {editor} {edited by\ \bibinfo {editor} {\bibfnamefont {W.}~\bibnamefont
  {Andreoni}}\ and\ \bibinfo {editor} {\bibfnamefont {S.}~\bibnamefont {Yip}}}\
  (\bibinfo  {publisher} {Springer International Publishing},\ \bibinfo
  {address} {Cham},\ \bibinfo {year} {2020})\ pp.\ \bibinfo {pages}
  {367--395}\BibitemShut {NoStop}%
\bibitem [{\citenamefont {Sangalli}\ and\ \citenamefont
  {Marini}(2015)}]{sangalli_ultra-fast_2015}%
  \BibitemOpen
  \bibfield  {author} {\bibinfo {author} {\bibfnamefont {D.}~\bibnamefont
  {Sangalli}}\ and\ \bibinfo {author} {\bibfnamefont {A.}~\bibnamefont
  {Marini}},\ }\href {\doibase 10.1209/0295-5075/110/47004} {\bibfield
  {journal} {\bibinfo  {journal} {Eurphys. Lett.}\ }\textbf {\bibinfo {volume}
  {110}},\ \bibinfo {pages} {47004} (\bibinfo {year} {2015})}\BibitemShut
  {NoStop}%
\bibitem [{\citenamefont {Molina-S\'{a}nchez}\ \emph
  {et~al.}(2017)\citenamefont {Molina-S\'{a}nchez}, \citenamefont {Sangalli},
  \citenamefont {Wirtz},\ and\ \citenamefont
  {Marini}}]{molina-sanchez_ab_2017}%
  \BibitemOpen
  \bibfield  {author} {\bibinfo {author} {\bibfnamefont {A.}~\bibnamefont
  {Molina-S\'{a}nchez}}, \bibinfo {author} {\bibfnamefont {D.}~\bibnamefont
  {Sangalli}}, \bibinfo {author} {\bibfnamefont {L.}~\bibnamefont {Wirtz}}, \
  and\ \bibinfo {author} {\bibfnamefont {A.}~\bibnamefont {Marini}},\ }\href
  {\doibase 10.1021/acs.nanolett.7b00175} {\bibfield  {journal} {\bibinfo
  {journal} {Nano Lett.}\ }\textbf {\bibinfo {volume} {17}},\ \bibinfo {pages}
  {4549} (\bibinfo {year} {2017})}\BibitemShut {NoStop}%
\bibitem [{\citenamefont {Selig}\ \emph {et~al.}(2016)\citenamefont {Selig},
  \citenamefont {Bergh\"{a}user}, \citenamefont {Raja}, \citenamefont {Nagler},
  \citenamefont {Sch\"{u}ller}, \citenamefont {Heinz}, \citenamefont {Korn},
  \citenamefont {Chernikov}, \citenamefont {Malic},\ and\ \citenamefont
  {Knorr}}]{selig_excitonic_2016}%
  \BibitemOpen
  \bibfield  {author} {\bibinfo {author} {\bibfnamefont {M.}~\bibnamefont
  {Selig}}, \bibinfo {author} {\bibfnamefont {G.}~\bibnamefont
  {Bergh\"{a}user}}, \bibinfo {author} {\bibfnamefont {A.}~\bibnamefont
  {Raja}}, \bibinfo {author} {\bibfnamefont {P.}~\bibnamefont {Nagler}},
  \bibinfo {author} {\bibfnamefont {C.}~\bibnamefont {Sch\"{u}ller}}, \bibinfo
  {author} {\bibfnamefont {T.~F.}\ \bibnamefont {Heinz}}, \bibinfo {author}
  {\bibfnamefont {T.}~\bibnamefont {Korn}}, \bibinfo {author} {\bibfnamefont
  {A.}~\bibnamefont {Chernikov}}, \bibinfo {author} {\bibfnamefont
  {E.}~\bibnamefont {Malic}}, \ and\ \bibinfo {author} {\bibfnamefont
  {A.}~\bibnamefont {Knorr}},\ }\href {\doibase 10.1038/ncomms13279} {\bibfield
   {journal} {\bibinfo  {journal} {Nat. Commun.}\ }\textbf {\bibinfo {volume}
  {7}},\ \bibinfo {pages} {13279} (\bibinfo {year} {2016})}\BibitemShut
  {NoStop}%
\bibitem [{\citenamefont {Trovatello}\ \emph {et~al.}(2020)\citenamefont
  {Trovatello}, \citenamefont {Miranda}, \citenamefont {Molina-S\'{a}nchez},
  \citenamefont {Borrego-Varillas}, \citenamefont {Manzoni}, \citenamefont
  {Moretti}, \citenamefont {Ganzer}, \citenamefont {Maiuri}, \citenamefont
  {Wang}, \citenamefont {Dumcenco}, \citenamefont {Kis}, \citenamefont {Wirtz},
  \citenamefont {Marini}, \citenamefont {Soavi}, \citenamefont {Ferrari},
  \citenamefont {Cerullo}, \citenamefont {Sangalli},\ and\ \citenamefont
  {Conte}}]{trovatello_strongly_2020}%
  \BibitemOpen
  \bibfield  {author} {\bibinfo {author} {\bibfnamefont {C.}~\bibnamefont
  {Trovatello}}, \bibinfo {author} {\bibfnamefont {H.~P.~C.}\ \bibnamefont
  {Miranda}}, \bibinfo {author} {\bibfnamefont {A.}~\bibnamefont
  {Molina-S\'{a}nchez}}, \bibinfo {author} {\bibfnamefont {R.}~\bibnamefont
  {Borrego-Varillas}}, \bibinfo {author} {\bibfnamefont {C.}~\bibnamefont
  {Manzoni}}, \bibinfo {author} {\bibfnamefont {L.}~\bibnamefont {Moretti}},
  \bibinfo {author} {\bibfnamefont {L.}~\bibnamefont {Ganzer}}, \bibinfo
  {author} {\bibfnamefont {M.}~\bibnamefont {Maiuri}}, \bibinfo {author}
  {\bibfnamefont {J.}~\bibnamefont {Wang}}, \bibinfo {author} {\bibfnamefont
  {D.}~\bibnamefont {Dumcenco}}, \bibinfo {author} {\bibfnamefont
  {A.}~\bibnamefont {Kis}}, \bibinfo {author} {\bibfnamefont {L.}~\bibnamefont
  {Wirtz}}, \bibinfo {author} {\bibfnamefont {A.}~\bibnamefont {Marini}},
  \bibinfo {author} {\bibfnamefont {G.}~\bibnamefont {Soavi}}, \bibinfo
  {author} {\bibfnamefont {A.~C.}\ \bibnamefont {Ferrari}}, \bibinfo {author}
  {\bibfnamefont {G.}~\bibnamefont {Cerullo}}, \bibinfo {author} {\bibfnamefont
  {D.}~\bibnamefont {Sangalli}}, \ and\ \bibinfo {author} {\bibfnamefont
  {S.~D.}\ \bibnamefont {Conte}},\ }\href {\doibase 10.1021/acsnano.0c00309}
  {\bibfield  {journal} {\bibinfo  {journal} {ACS Nano}\ }\textbf {\bibinfo
  {volume} {14}},\ \bibinfo {pages} {5700} (\bibinfo {year}
  {2020})}\BibitemShut {NoStop}%
\bibitem [{\citenamefont {Flick}\ \emph {et~al.}(2018)\citenamefont {Flick},
  \citenamefont {Sch\"{a}fer}, \citenamefont {Ruggenthaler}, \citenamefont
  {Appel},\ and\ \citenamefont {Rubio}}]{flick_ab_2018}%
  \BibitemOpen
  \bibfield  {author} {\bibinfo {author} {\bibfnamefont {J.}~\bibnamefont
  {Flick}}, \bibinfo {author} {\bibfnamefont {C.}~\bibnamefont {Sch\"{a}fer}},
  \bibinfo {author} {\bibfnamefont {M.}~\bibnamefont {Ruggenthaler}}, \bibinfo
  {author} {\bibfnamefont {H.}~\bibnamefont {Appel}}, \ and\ \bibinfo {author}
  {\bibfnamefont {A.}~\bibnamefont {Rubio}},\ }\href {\doibase
  10.1021/acsphotonics.7b01279} {\bibfield  {journal} {\bibinfo  {journal} {ACS
  Photonics}\ }\textbf {\bibinfo {volume} {5}},\ \bibinfo {pages} {992}
  (\bibinfo {year} {2018})}\BibitemShut {NoStop}%
\bibitem [{\citenamefont {Ojambati}\ \emph {et~al.}(2019)\citenamefont
  {Ojambati}, \citenamefont {Chikkaraddy}, \citenamefont {Deacon},
  \citenamefont {Horton}, \citenamefont {Kos}, \citenamefont {Turek},
  \citenamefont {Keyser},\ and\ \citenamefont
  {Baumberg}}]{ojambati_quantum_2019}%
  \BibitemOpen
  \bibfield  {author} {\bibinfo {author} {\bibfnamefont {O.~S.}\ \bibnamefont
  {Ojambati}}, \bibinfo {author} {\bibfnamefont {R.}~\bibnamefont
  {Chikkaraddy}}, \bibinfo {author} {\bibfnamefont {W.~D.}\ \bibnamefont
  {Deacon}}, \bibinfo {author} {\bibfnamefont {M.}~\bibnamefont {Horton}},
  \bibinfo {author} {\bibfnamefont {D.}~\bibnamefont {Kos}}, \bibinfo {author}
  {\bibfnamefont {V.~A.}\ \bibnamefont {Turek}}, \bibinfo {author}
  {\bibfnamefont {U.~F.}\ \bibnamefont {Keyser}}, \ and\ \bibinfo {author}
  {\bibfnamefont {J.~J.}\ \bibnamefont {Baumberg}},\ }\href {\doibase
  10.1038/s41467-019-08611-5} {\bibfield  {journal} {\bibinfo  {journal} {Nat.
  Commun.}\ }\textbf {\bibinfo {volume} {10}},\ \bibinfo {pages} {1049}
  (\bibinfo {year} {2019})}\BibitemShut {NoStop}%
\bibitem [{\citenamefont {Sch\"{a}fer}\ \emph {et~al.}(2019)\citenamefont
  {Sch\"{a}fer}, \citenamefont {Ruggenthaler}, \citenamefont {Appel},\ and\
  \citenamefont {Rubio}}]{schafer_modification_2019}%
  \BibitemOpen
  \bibfield  {author} {\bibinfo {author} {\bibfnamefont {C.}~\bibnamefont
  {Sch\"{a}fer}}, \bibinfo {author} {\bibfnamefont {M.}~\bibnamefont
  {Ruggenthaler}}, \bibinfo {author} {\bibfnamefont {H.}~\bibnamefont {Appel}},
  \ and\ \bibinfo {author} {\bibfnamefont {A.}~\bibnamefont {Rubio}},\ }\href
  {\doibase 10.1073/pnas.1814178116} {\bibfield  {journal} {\bibinfo  {journal}
  {PNAS}\ }\textbf {\bibinfo {volume} {116}},\ \bibinfo {pages} {4883}
  (\bibinfo {year} {2019})}\BibitemShut {NoStop}%
\bibitem [{\citenamefont {Li}\ \emph {et~al.}(2021)\citenamefont {Li},
  \citenamefont {Mandal},\ and\ \citenamefont {Huo}}]{li_cavity_2021}%
  \BibitemOpen
  \bibfield  {author} {\bibinfo {author} {\bibfnamefont {X.}~\bibnamefont
  {Li}}, \bibinfo {author} {\bibfnamefont {A.}~\bibnamefont {Mandal}}, \ and\
  \bibinfo {author} {\bibfnamefont {P.}~\bibnamefont {Huo}},\ }\href {\doibase
  10.1038/s41467-021-21610-9} {\bibfield  {journal} {\bibinfo  {journal} {Nat.
  Commun.}\ }\textbf {\bibinfo {volume} {12}},\ \bibinfo {pages} {1315}
  (\bibinfo {year} {2021})}\BibitemShut {NoStop}%
\bibitem [{\citenamefont {Karlsson}\ \emph {et~al.}(2018)\citenamefont
  {Karlsson}, \citenamefont {van Leeuwen}, \citenamefont {Perfetto},\ and\
  \citenamefont {Stefanucci}}]{karlsson_generalized_2018}%
  \BibitemOpen
  \bibfield  {author} {\bibinfo {author} {\bibfnamefont {D.}~\bibnamefont
  {Karlsson}}, \bibinfo {author} {\bibfnamefont {R.}~\bibnamefont {van
  Leeuwen}}, \bibinfo {author} {\bibfnamefont {E.}~\bibnamefont {Perfetto}}, \
  and\ \bibinfo {author} {\bibfnamefont {G.}~\bibnamefont {Stefanucci}},\
  }\href {\doibase 10.1103/PhysRevB.98.115148} {\bibfield  {journal} {\bibinfo
  {journal} {Phys. Rev. B}\ }\textbf {\bibinfo {volume} {98}},\ \bibinfo
  {pages} {115148} (\bibinfo {year} {2018})}\BibitemShut {NoStop}%
\bibitem [{\citenamefont {Bostr\"{o}m}\ \emph {et~al.}(2018)\citenamefont
  {Bostr\"{o}m}, \citenamefont {Mikkelsen}, \citenamefont {Verdozzi},
  \citenamefont {Perfetto},\ and\ \citenamefont
  {Stefanucci}}]{bostrom_charge_2018}%
  \BibitemOpen
  \bibfield  {author} {\bibinfo {author} {\bibfnamefont {E.~V.~n.}\
  \bibnamefont {Bostr\"{o}m}}, \bibinfo {author} {\bibfnamefont
  {A.}~\bibnamefont {Mikkelsen}}, \bibinfo {author} {\bibfnamefont
  {C.}~\bibnamefont {Verdozzi}}, \bibinfo {author} {\bibfnamefont
  {E.}~\bibnamefont {Perfetto}}, \ and\ \bibinfo {author} {\bibfnamefont
  {G.}~\bibnamefont {Stefanucci}},\ }\href {\doibase
  10.1021/acs.nanolett.7b03995} {\bibfield  {journal} {\bibinfo  {journal}
  {Nano Lett.}\ }\textbf {\bibinfo {volume} {18}},\ \bibinfo {pages} {785}
  (\bibinfo {year} {2018})}\BibitemShut {NoStop}%
\bibitem [{\citenamefont {Perfetto}\ \emph {et~al.}(2018)\citenamefont
  {Perfetto}, \citenamefont {Sangalli}, \citenamefont {Marini},\ and\
  \citenamefont {Stefanucci}}]{perfetto_ultrafast_2018}%
  \BibitemOpen
  \bibfield  {author} {\bibinfo {author} {\bibfnamefont {E.}~\bibnamefont
  {Perfetto}}, \bibinfo {author} {\bibfnamefont {D.}~\bibnamefont {Sangalli}},
  \bibinfo {author} {\bibfnamefont {A.}~\bibnamefont {Marini}}, \ and\ \bibinfo
  {author} {\bibfnamefont {G.}~\bibnamefont {Stefanucci}},\ }\href {\doibase
  10.1021/acs.jpclett.8b00025} {\bibfield  {journal} {\bibinfo  {journal} {J.
  Phys. Chem. Lett.}\ }\textbf {\bibinfo {volume} {9}},\ \bibinfo {pages}
  {1353} (\bibinfo {year} {2018})}\BibitemShut {NoStop}%
\bibitem [{\citenamefont {Murakami}\ \emph {et~al.}(2017)\citenamefont
  {Murakami}, \citenamefont {Tsuji}, \citenamefont {Eckstein},\ and\
  \citenamefont {Werner}}]{murakami_nonequilibrium_2017}%
  \BibitemOpen
  \bibfield  {author} {\bibinfo {author} {\bibfnamefont {Y.}~\bibnamefont
  {Murakami}}, \bibinfo {author} {\bibfnamefont {N.}~\bibnamefont {Tsuji}},
  \bibinfo {author} {\bibfnamefont {M.}~\bibnamefont {Eckstein}}, \ and\
  \bibinfo {author} {\bibfnamefont {P.}~\bibnamefont {Werner}},\ }\href
  {\doibase 10.1103/PhysRevB.96.045125} {\bibfield  {journal} {\bibinfo
  {journal} {Phys. Rev. B}\ }\textbf {\bibinfo {volume} {96}},\ \bibinfo
  {pages} {045125} (\bibinfo {year} {2017})}\BibitemShut {NoStop}%
\bibitem [{\citenamefont {Fan}(1951)}]{fan_temperature_1951}%
  \BibitemOpen
  \bibfield  {author} {\bibinfo {author} {\bibfnamefont {H.~Y.}\ \bibnamefont
  {Fan}},\ }\href {\doibase 10.1103/PhysRev.82.900} {\bibfield  {journal}
  {\bibinfo  {journal} {Phys. Rev.}\ }\textbf {\bibinfo {volume} {82}},\
  \bibinfo {pages} {900} (\bibinfo {year} {1951})}\BibitemShut {NoStop}%
\bibitem [{\citenamefont {Murakami}\ \emph {et~al.}(2016)\citenamefont
  {Murakami}, \citenamefont {Werner}, \citenamefont {Tsuji},\ and\
  \citenamefont {Aoki}}]{murakami_multiple_2016}%
  \BibitemOpen
  \bibfield  {author} {\bibinfo {author} {\bibfnamefont {Y.}~\bibnamefont
  {Murakami}}, \bibinfo {author} {\bibfnamefont {P.}~\bibnamefont {Werner}},
  \bibinfo {author} {\bibfnamefont {N.}~\bibnamefont {Tsuji}}, \ and\ \bibinfo
  {author} {\bibfnamefont {H.}~\bibnamefont {Aoki}},\ }\href {\doibase
  10.1103/PhysRevB.93.094509} {\bibfield  {journal} {\bibinfo  {journal} {Phys.
  Rev. B}\ }\textbf {\bibinfo {volume} {93}},\ \bibinfo {pages} {094509}
  (\bibinfo {year} {2016})}\BibitemShut {NoStop}%
\bibitem [{\citenamefont {Giustino}\ \emph {et~al.}(2007)\citenamefont
  {Giustino}, \citenamefont {Cohen},\ and\ \citenamefont
  {Louie}}]{giustino_electron-phonon_2007}%
  \BibitemOpen
  \bibfield  {author} {\bibinfo {author} {\bibfnamefont {F.}~\bibnamefont
  {Giustino}}, \bibinfo {author} {\bibfnamefont {M.~L.}\ \bibnamefont {Cohen}},
  \ and\ \bibinfo {author} {\bibfnamefont {S.~G.}\ \bibnamefont {Louie}},\
  }\href {\doibase 10.1103/PhysRevB.76.165108} {\bibfield  {journal} {\bibinfo
  {journal} {Phys. Rev. B}\ }\textbf {\bibinfo {volume} {76}},\ \bibinfo
  {pages} {165108} (\bibinfo {year} {2007})}\BibitemShut {NoStop}%
\bibitem [{\citenamefont {Marini}\ \emph {et~al.}(2015)\citenamefont {Marini},
  \citenamefont {Ponc\'{e}},\ and\ \citenamefont
  {Gonze}}]{marini_many-body_2015}%
  \BibitemOpen
  \bibfield  {author} {\bibinfo {author} {\bibfnamefont {A.}~\bibnamefont
  {Marini}}, \bibinfo {author} {\bibfnamefont {S.}~\bibnamefont {Ponc\'{e}}}, \
  and\ \bibinfo {author} {\bibfnamefont {X.}~\bibnamefont {Gonze}},\ }\href
  {\doibase 10.1103/PhysRevB.91.224310} {\bibfield  {journal} {\bibinfo
  {journal} {Phys. Rev. B}\ }\textbf {\bibinfo {volume} {91}},\ \bibinfo
  {pages} {224310} (\bibinfo {year} {2015})}\BibitemShut {NoStop}%
\bibitem [{\citenamefont {Restrepo}\ \emph {et~al.}(2009)\citenamefont
  {Restrepo}, \citenamefont {Varga},\ and\ \citenamefont
  {Pantelides}}]{restrepo_first-principles_2009}%
  \BibitemOpen
  \bibfield  {author} {\bibinfo {author} {\bibfnamefont {O.~D.}\ \bibnamefont
  {Restrepo}}, \bibinfo {author} {\bibfnamefont {K.}~\bibnamefont {Varga}}, \
  and\ \bibinfo {author} {\bibfnamefont {S.~T.}\ \bibnamefont {Pantelides}},\
  }\href {\doibase 10.1063/1.3147189} {\bibfield  {journal} {\bibinfo
  {journal} {Appl. Phys. Lett.}\ }\textbf {\bibinfo {volume} {94}},\ \bibinfo
  {pages} {212103} (\bibinfo {year} {2009})}\BibitemShut {NoStop}%
\bibitem [{\citenamefont {Verdi}\ \emph {et~al.}(2017)\citenamefont {Verdi},
  \citenamefont {Caruso},\ and\ \citenamefont {Giustino}}]{verdi_origin_2017}%
  \BibitemOpen
  \bibfield  {author} {\bibinfo {author} {\bibfnamefont {C.}~\bibnamefont
  {Verdi}}, \bibinfo {author} {\bibfnamefont {F.}~\bibnamefont {Caruso}}, \
  and\ \bibinfo {author} {\bibfnamefont {F.}~\bibnamefont {Giustino}},\ }\href
  {\doibase 10.1038/ncomms15769} {\bibfield  {journal} {\bibinfo  {journal}
  {Nat. Commun.}\ }\textbf {\bibinfo {volume} {8}},\ \bibinfo {pages} {15769}
  (\bibinfo {year} {2017})}\BibitemShut {NoStop}%
\bibitem [{\citenamefont {Mahan}(2000)}]{mahan_many-particle_2000}%
  \BibitemOpen
  \bibfield  {author} {\bibinfo {author} {\bibfnamefont {G.}~\bibnamefont
  {Mahan}},\ }\href {\doibase 10.1007/978-1-4757-5714-9} {\emph {\bibinfo
  {title} {Many-particle physics}}},\ \bibinfo {edition} {3rd}\ ed.\ (\bibinfo
  {publisher} {Springer},\ \bibinfo {address} {US, New York},\ \bibinfo {year}
  {2000})\BibitemShut {NoStop}%
\bibitem [{\citenamefont {Giustino}(2017)}]{giustino_electron-phonon_2017}%
  \BibitemOpen
  \bibfield  {author} {\bibinfo {author} {\bibfnamefont {F.}~\bibnamefont
  {Giustino}},\ }\href {\doibase 10.1103/RevModPhys.89.015003} {\bibfield
  {journal} {\bibinfo  {journal} {Rev. Mod. Phys.}\ }\textbf {\bibinfo {volume}
  {89}},\ \bibinfo {pages} {015003} (\bibinfo {year} {2017})}\BibitemShut
  {NoStop}%
\bibitem [{\citenamefont {Caruso}\ \emph {et~al.}(2017)\citenamefont {Caruso},
  \citenamefont {Hoesch}, \citenamefont {Achatz}, \citenamefont {Serrano},
  \citenamefont {Krisch}, \citenamefont {Bustarret},\ and\ \citenamefont
  {Giustino}}]{caruso_nonadiabatic_2017}%
  \BibitemOpen
  \bibfield  {author} {\bibinfo {author} {\bibfnamefont {F.}~\bibnamefont
  {Caruso}}, \bibinfo {author} {\bibfnamefont {M.}~\bibnamefont {Hoesch}},
  \bibinfo {author} {\bibfnamefont {P.}~\bibnamefont {Achatz}}, \bibinfo
  {author} {\bibfnamefont {J.}~\bibnamefont {Serrano}}, \bibinfo {author}
  {\bibfnamefont {M.}~\bibnamefont {Krisch}}, \bibinfo {author} {\bibfnamefont
  {E.}~\bibnamefont {Bustarret}}, \ and\ \bibinfo {author} {\bibfnamefont
  {F.}~\bibnamefont {Giustino}},\ }\href {\doibase
  10.1103/PhysRevLett.119.017001} {\bibfield  {journal} {\bibinfo  {journal}
  {Phys. Rev. Lett.}\ }\textbf {\bibinfo {volume} {119}},\ \bibinfo {pages}
  {017001} (\bibinfo {year} {2017})}\BibitemShut {NoStop}%
\bibitem [{\citenamefont {Kuleff}\ \emph {et~al.}(2005)\citenamefont {Kuleff},
  \citenamefont {Breidbach},\ and\ \citenamefont
  {Cederbaum}}]{kuleff_multielectron_2005}%
  \BibitemOpen
  \bibfield  {author} {\bibinfo {author} {\bibfnamefont {A.~I.}\ \bibnamefont
  {Kuleff}}, \bibinfo {author} {\bibfnamefont {J.}~\bibnamefont {Breidbach}}, \
  and\ \bibinfo {author} {\bibfnamefont {L.~S.}\ \bibnamefont {Cederbaum}},\
  }\href {\doibase 10.1063/1.1961341} {\bibfield  {journal} {\bibinfo
  {journal} {J. Chem. Phys.}\ }\textbf {\bibinfo {volume} {123}},\ \bibinfo
  {pages} {044111} (\bibinfo {year} {2005})}\BibitemShut {NoStop}%
\bibitem [{\citenamefont {Kuleff}\ and\ \citenamefont
  {Cederbaum}(2007)}]{kuleff_charge_2007}%
  \BibitemOpen
  \bibfield  {author} {\bibinfo {author} {\bibfnamefont {A.~I.}\ \bibnamefont
  {Kuleff}}\ and\ \bibinfo {author} {\bibfnamefont {L.~S.}\ \bibnamefont
  {Cederbaum}},\ }\href {\doibase 10.1016/j.chemphys.2007.04.012} {\bibfield
  {journal} {\bibinfo  {journal} {Chem. Phys.}\ }\textbf {\bibinfo {volume}
  {338}},\ \bibinfo {pages} {320} (\bibinfo {year} {2007})}\BibitemShut
  {NoStop}%
\bibitem [{\citenamefont {Cooper}\ and\ \citenamefont
  {Averbukh}(2013)}]{cooper_single-photon_2013}%
  \BibitemOpen
  \bibfield  {author} {\bibinfo {author} {\bibfnamefont {B.}~\bibnamefont
  {Cooper}}\ and\ \bibinfo {author} {\bibfnamefont {V.}~\bibnamefont
  {Averbukh}},\ }\href {\doibase 10.1103/PhysRevLett.111.083004} {\bibfield
  {journal} {\bibinfo  {journal} {Phys. Rev. Lett.}\ }\textbf {\bibinfo
  {volume} {111}},\ \bibinfo {pages} {083004} (\bibinfo {year}
  {2013})}\BibitemShut {NoStop}%
\bibitem [{\citenamefont {Perfetto}\ \emph {et~al.}(2019)\citenamefont
  {Perfetto}, \citenamefont {Sangalli}, \citenamefont {Palummo}, \citenamefont
  {Marini},\ and\ \citenamefont {Stefanucci}}]{perfetto_first-principles_2019}%
  \BibitemOpen
  \bibfield  {author} {\bibinfo {author} {\bibfnamefont {E.}~\bibnamefont
  {Perfetto}}, \bibinfo {author} {\bibfnamefont {D.}~\bibnamefont {Sangalli}},
  \bibinfo {author} {\bibfnamefont {M.}~\bibnamefont {Palummo}}, \bibinfo
  {author} {\bibfnamefont {A.}~\bibnamefont {Marini}}, \ and\ \bibinfo {author}
  {\bibfnamefont {G.}~\bibnamefont {Stefanucci}},\ }\href {\doibase
  10.1021/acs.jctc.9b00170} {\bibfield  {journal} {\bibinfo  {journal} {J.
  Chem. Theory Comput.}\ }\textbf {\bibinfo {volume} {15}},\ \bibinfo {pages}
  {4526} (\bibinfo {year} {2019})}\BibitemShut {NoStop}%
\bibitem [{\citenamefont {Perfetto}\ and\ \citenamefont
  {Stefanucci}(2018)}]{perfetto_cheers:_2018}%
  \BibitemOpen
  \bibfield  {author} {\bibinfo {author} {\bibfnamefont {E.}~\bibnamefont
  {Perfetto}}\ and\ \bibinfo {author} {\bibfnamefont {G.}~\bibnamefont
  {Stefanucci}},\ }\href {\doibase 10.1088/1361-648X/aae675} {\bibfield
  {journal} {\bibinfo  {journal} {J. Phys. Condens. Matter}\ }\textbf {\bibinfo
  {volume} {30}},\ \bibinfo {pages} {465901} (\bibinfo {year}
  {2018})}\BibitemShut {NoStop}%
\bibitem [{\citenamefont {Yang}\ \emph {et~al.}(2021)\citenamefont {Yang},
  \citenamefont {Ou}, \citenamefont {Pei}, \citenamefont {Wang}, \citenamefont
  {Weng}, \citenamefont {Shuai}, \citenamefont {Mullen},\ and\ \citenamefont
  {Shao}}]{yang_quantum-electrodynamical_2021}%
  \BibitemOpen
  \bibfield  {author} {\bibinfo {author} {\bibfnamefont {J.}~\bibnamefont
  {Yang}}, \bibinfo {author} {\bibfnamefont {Q.}~\bibnamefont {Ou}}, \bibinfo
  {author} {\bibfnamefont {Z.}~\bibnamefont {Pei}}, \bibinfo {author}
  {\bibfnamefont {H.}~\bibnamefont {Wang}}, \bibinfo {author} {\bibfnamefont
  {B.}~\bibnamefont {Weng}}, \bibinfo {author} {\bibfnamefont {Z.}~\bibnamefont
  {Shuai}}, \bibinfo {author} {\bibfnamefont {K.}~\bibnamefont {Mullen}}, \
  and\ \bibinfo {author} {\bibfnamefont {Y.}~\bibnamefont {Shao}},\ }\href
  {\doibase 10.1063/5.0057542} {\bibfield  {journal} {\bibinfo  {journal} {J.
  Chem. Phys.}\ }\textbf {\bibinfo {volume} {155}},\ \bibinfo {pages} {064107}
  (\bibinfo {year} {2021})}\BibitemShut {NoStop}%
\bibitem [{\citenamefont {Autler}\ and\ \citenamefont
  {Townes}(1955)}]{autler_stark_1955}%
  \BibitemOpen
  \bibfield  {author} {\bibinfo {author} {\bibfnamefont {S.~H.}\ \bibnamefont
  {Autler}}\ and\ \bibinfo {author} {\bibfnamefont {C.~H.}\ \bibnamefont
  {Townes}},\ }\href {\doibase 10.1103/PhysRev.100.703} {\bibfield  {journal}
  {\bibinfo  {journal} {Phys. Rev.}\ }\textbf {\bibinfo {volume} {100}},\
  \bibinfo {pages} {703} (\bibinfo {year} {1955})}\BibitemShut {NoStop}%
\bibitem [{\citenamefont {Perfetto}\ and\ \citenamefont
  {Stefanucci}(2015)}]{perfetto_exact_2015}%
  \BibitemOpen
  \bibfield  {author} {\bibinfo {author} {\bibfnamefont {E.}~\bibnamefont
  {Perfetto}}\ and\ \bibinfo {author} {\bibfnamefont {G.}~\bibnamefont
  {Stefanucci}},\ }\href {\doibase 10.1103/PhysRevA.91.033416} {\bibfield
  {journal} {\bibinfo  {journal} {Phys. Rev. A}\ }\textbf {\bibinfo {volume}
  {91}},\ \bibinfo {pages} {033416} (\bibinfo {year} {2015})}\BibitemShut
  {NoStop}%
\bibitem [{\citenamefont {Sangalli}\ \emph {et~al.}(2019)\citenamefont
  {Sangalli}, \citenamefont {Ferretti}, \citenamefont {Miranda}, \citenamefont
  {Attaccalite}, \citenamefont {Marri}, \citenamefont {Cannuccia},
  \citenamefont {Melo}, \citenamefont {Marsili}, \citenamefont {Paleari},
  \citenamefont {Marrazzo}, \citenamefont {Prandini}, \citenamefont
  {Bonf\`{a}}, \citenamefont {Atambo}, \citenamefont {Affinito}, \citenamefont
  {Palummo}, \citenamefont {Molina-S\'{a}nchez}, \citenamefont {Hogan},
  \citenamefont {Gr\"{u}ning}, \citenamefont {Varsano},\ and\ \citenamefont
  {Marini}}]{sangalli_many-body_2019}%
  \BibitemOpen
  \bibfield  {author} {\bibinfo {author} {\bibfnamefont {D.}~\bibnamefont
  {Sangalli}}, \bibinfo {author} {\bibfnamefont {A.}~\bibnamefont {Ferretti}},
  \bibinfo {author} {\bibfnamefont {H.}~\bibnamefont {Miranda}}, \bibinfo
  {author} {\bibfnamefont {C.}~\bibnamefont {Attaccalite}}, \bibinfo {author}
  {\bibfnamefont {I.}~\bibnamefont {Marri}}, \bibinfo {author} {\bibfnamefont
  {E.}~\bibnamefont {Cannuccia}}, \bibinfo {author} {\bibfnamefont
  {P.}~\bibnamefont {Melo}}, \bibinfo {author} {\bibfnamefont {M.}~\bibnamefont
  {Marsili}}, \bibinfo {author} {\bibfnamefont {F.}~\bibnamefont {Paleari}},
  \bibinfo {author} {\bibfnamefont {A.}~\bibnamefont {Marrazzo}}, \bibinfo
  {author} {\bibfnamefont {G.}~\bibnamefont {Prandini}}, \bibinfo {author}
  {\bibfnamefont {P.}~\bibnamefont {Bonf\`{a}}}, \bibinfo {author}
  {\bibfnamefont {M.~O.}\ \bibnamefont {Atambo}}, \bibinfo {author}
  {\bibfnamefont {F.}~\bibnamefont {Affinito}}, \bibinfo {author}
  {\bibfnamefont {M.}~\bibnamefont {Palummo}}, \bibinfo {author} {\bibfnamefont
  {A.}~\bibnamefont {Molina-S\'{a}nchez}}, \bibinfo {author} {\bibfnamefont
  {C.}~\bibnamefont {Hogan}}, \bibinfo {author} {\bibfnamefont
  {M.}~\bibnamefont {Gr\"{u}ning}}, \bibinfo {author} {\bibfnamefont
  {D.}~\bibnamefont {Varsano}}, \ and\ \bibinfo {author} {\bibfnamefont
  {A.}~\bibnamefont {Marini}},\ }\href {\doibase 10.1088/1361-648X/ab15d0}
  {\bibfield  {journal} {\bibinfo  {journal} {J. Phys. Condens. Matter}\
  }\textbf {\bibinfo {volume} {31}},\ \bibinfo {pages} {325902} (\bibinfo
  {year} {2019})}\BibitemShut {NoStop}%
\bibitem [{\citenamefont {Reining}(2016)}]{pavarini_linear_2016}%
  \BibitemOpen
  \bibfield  {author} {\bibinfo {author} {\bibfnamefont {L.}~\bibnamefont
  {Reining}},\ }in\ \href@noop {} {\emph {\bibinfo {booktitle} {Quantum
  materials: experiments and theory: lecture notes of the {Autumn} {School} on
  {Correlated} {Electrons} 2016}}},\ \bibinfo {series} {Modeling and
  {Simulation}}, Vol.~\bibinfo {volume} {6},\ \bibinfo {editor} {edited by\
  \bibinfo {editor} {\bibfnamefont {E.}~\bibnamefont {Pavarini}}, \bibinfo
  {editor} {\bibfnamefont {E.}~\bibnamefont {Koch}}, \bibinfo {editor}
  {\bibfnamefont {J.}~\bibnamefont {van~den Brink}}, \ and\ \bibinfo {editor}
  {\bibfnamefont {G.}~\bibnamefont {Sawatzky}}}\ (\bibinfo  {publisher}
  {Forschungszentrum J\"{u}lich GmbH, Institute for Advanced Simulation},\
  \bibinfo {year} {2016})\BibitemShut {NoStop}%
\end{thebibliography}


\end{document}